\documentclass[usenatbib]{mnras}

\usepackage{multirow}
\usepackage{rotating}
\usepackage{colortbl}
\usepackage{color}
\usepackage[fleqn]{amsmath}
\usepackage{amssymb}
\usepackage{amsfonts}
\usepackage{verbatim}
\usepackage{scalefnt}
\usepackage[percent]{overpic}
\usepackage[T1]{fontenc} 
\usepackage{aecompl} 
\usepackage{ulem}
\usepackage{cleveref}
\usepackage{numprint}

\usepackage{float}

\def\lsim{\mathrel{\rlap{\lower 3pt \hbox{$\sim$}} \raise 2.0pt \hbox{$<$}}}
\def\gsim{\mathrel{\rlap{\lower 3pt \hbox{$\sim$}} \raise 2.0pt \hbox{$>$}}}

\newcommand{\comments}[1]{} 

\newcommand\T{\rule{0pt}{2.8ex}}       
\newcommand\B{\rule[-1.4ex]{0pt}{0pt}} 

\title[Improved gravitational radiation time-scales II]{Improved gravitational radiation time-scales II: spin-orbit contributions and environmental perturbations}

\author[L. Zwick et al.]{Lorenz Zwick,$^{1}$\thanks{E-mail: zwicklo@ics.uzh.ch}
Pedro~R. Capelo,$^{1}$
Elisa Bortolas,$^{2,1}$\newauthor
Ver\'{o}nica V\'{a}zquez-Aceves,$^{3}$
Lucio Mayer$^{1}$ and
Pau Amaro-Seoane$^{3,4,5,6,7}$
\\
$^{1}$Center for Theoretical Astrophysics and Cosmology, Institute for Computational Science, University of Zurich,\\ Winterthurerstrasse 190, CH-8057 Z{\"u}rich, Switzerland\\
$^{2}$Dipartimento di Fisica ``G. Occhialini'', Università degli Studi di Milano-Bicocca, Piazza della Scienza 3, I-20126 Milano, Italy\\
$^{3}$Institute of Applied Mathematics, Academy of Mathematics and Systems Science, Chinese Academy of Sciences, 100190 Beijing, China\\
$^{4}$Universitat Polit\`{e}cnica de Val\`{e}ncia, IGIC, Spain\\
$^{5}$Deutsches Elektronen Synchrotron DESY, Platanenallee 6, D-15738 Zeuthen, Germany\\
$^{6}$Kavli Institute for Astronomy and Astrophysics at Peking University, 100871 Beijing, China \\
$^{7}$Zentrum f{\"u}r Astronomie und Astrophysik, TU Berlin, Hardenbergstra{\ss}e 36, D-10623 Berlin, Germany}

\date{Accepted XXX. Received YYY; in original form ZZZ}

\pubyear{2021}

\begin{document}

\label{firstpage}

\pagerange{\pageref{firstpage}--\pageref{lastpage}}

\maketitle


\begin{abstract}
Peters' formula is an analytical estimate of the time-scale of gravitational wave (GW)-induced coalescence of binary systems. It is used in countless applications, where the convenience of a simple formula outweighs the need for precision. However, many promising sources of the Laser Interferometer Space Antenna (LISA), such as supermassive black hole binaries and extreme mass-ratio inspirals (EMRIs), are expected to enter the LISA band with highly eccentric ($e \gsim 0.9$) and highly relativistic orbits. These are exactly the two limits in which Peters' estimate performs the worst. In this work, we expand upon previous results and give simple analytical fits to quantify how the inspiral time-scale is affected by the relative 1.5 post-Newtonian (PN) hereditary fluxes and spin-orbit couplings. We discuss several cases that demand a more accurate GW time-scale. We show how this can have a major influence on quantities that are relevant for LISA event-rate estimates, such as the EMRI critical semi-major axis. We further discuss two types of environmental perturbations that can play a role in the inspiral phase: the gravitational interaction with a third massive body and the energy loss due to dynamical friction and torques from a surrounding gas medium ubiquitous in galactic nuclei. With the aid of PN corrections to the time-scale in vacuum, we find simple analytical expressions for the regions of phase space in which environmental perturbations are of comparable strength to the effects of any particular PN order, being able to qualitatively reproduce the results of much more sophisticated analyses.
\end{abstract}

\begin{keywords}
black hole physics -- gravitational waves -- methods: analytical.
\end{keywords}


\section{Introduction}\label{sec:introduction}

With the launch of space-borne gravitational-wave (GW) detectors in sight (e.g. the Laser Interferometer Space Antenna, LISA; \citealt{pau_2013,Barack_et_al_2019}), many questions still have to be answered with regards to the possible sources of signal. In particular, large uncertainties remain on the expected event rates for different types of sources and for different formation channels \citep[e.g.][]{klein,Babak}. An essential part of these estimates is a model of gravitational radiation and its effects on the orbital dynamics of a system. Very often, it is convenient to express this effect with a simple analytical time-scale.

Already in the early 1960s, \citet{Peters_Mathews_1963} calculated the radiation reaction equations and extracted a simple analytical formula for the time-scale of GW-induced decay of a binary. This formula, commonly known as Peters' time-scale, has since seen an incredible amount of use, and more recently has been employed in many event-rate estimates and other source-modelling applications \citep[see, e.g.][amongst many others]{2015MNRAS.447L..80F,2016ApJ...828...77V,2019MNRAS.485.2125B}. Peters' formula is thought to suffice as a first approximation, especially because it is often used in conjunction with other dynamical time-scales that are themselves only order-of-magnitude estimates \citep[see, e.g.][]{2015MNRAS.454L..66S}. However, many of LISA most promising sources, such as supermassive black hole (SMBH) binaries and extreme/intermediate/extremely large mass-ratio inspirals (EMRIs/IMRIs/XMRIs), are expected to exhibit highly eccentric and highly relativistic orbits \citep[see, e.g.][]{Amaro-Seoane2007,Antonini2016,Bonetti2016,Bonetti_et_al_2018,Khan_et_al_2018,pau1,Bonetti2019,Giacobbo2019,Giacobbo2019b,Amaro_Seoane_2019}, which are the two regimes in which Peters' estimate performs the worst. The commonly used formula does not account for the Newtonian evolution of the eccentricity. Moreover, it is based on the quadrupole radiation of a Keplerian binary and therefore fails to capture the more complex post-Newtonian (PN) behaviour of highly relativistic orbits.

In a recent paper \citep[hereafter Paper~I]{paper1}, we sought to fix these issues, in service of reducing the unnecessary errors that might arise from using Peters' formula in its common form. We produced an updated formula that takes into account the eccentricity evolution and the lowest-order PN correction, while still being algebraically very simple.

In this paper, we wish to extend our analysis to the next order of the PN series, for the two following reasons. Firstly, it will serve as a proof of good convergence for the previous results. Secondly, two qualitatively new effects emerge at the next order: hereditary (i.e. tail) and spin-orbit contributions. Hereditary terms in the PN flux equations are known to be disproportionately important in the amount of cycles that some GW sources complete in the LISA band \citep[see, e.g.][]{Blanchet2014}. Moreover, a significant fraction of SMBHs are expected to have large ($\gsim 0.9$) dimensionless spin values \citep[see, e.g.][]{Reynolds_2020}, whose effects are known to significantly alter the event rates of some GW sources, especially EMRIs \citep{pau_2013}. This, combined with the weak convergence of the PN series in the case of extreme mass ratios \citep[see, e.g.][]{2008PhRvD..77l4006Y,PhysRevD.84.024029}, suggests that there is still precision to be gained beyond the lowest-order PN correction.

At some point, however, one inevitably clashes against the fact that most GW sources do not evolve in a pure vacuum. Even at very small separations, the inspiral process can be perturbed by the presence of other forces and the vacuum time-scale can be a precise estimate only if these environmental perturbations are weak. Therefore, we take a look at two types of environmental perturbations to the inspiral time-scale. Firstly, we investigate the gravitational influence of a third body. Secondly, we compare the strength of gas friction and gas torques for an inspiral embedded in a gaseous disc. In both cases, we express the results as a series of characteristic orbital separations at which the environmental effects influence the inspiral time-scale as much as a particular PN order. Even though these results are only order-of-magnitude estimates, they clearly illustrate the regions of validity and usefulness for the corrections we propose, and more generally for gravitational radiation time-scales.

With the goals of this paper set, we wish to refer any reader not at least partly familiar with Peters' formula to the original work from 1963 and to Peters' Ph.D. thesis \citep{Peters_Mathews_1963,Peters_1964}, as well as to a general review of PN theory in \cite{Schaefer_rev}. We also refer to Paper~I for a more in-depth discussion of the limitations of Peters' derivation and on the considerations required to go beyond Keplerian orbits and quadrupolar radiation. Section~\ref{sec:methods} presents a brief but complete overview of the concepts that were used in this paper. In Section~\ref{sec:firstorder}, we present our derivation of the 1.5 PN correction factors to Peters' formula, and discuss how their inclusion can affect current event-rate estimates for EMRIs. In Section~\ref{sec:environment}, we discuss environmental effects and derive several useful characteristic radii. Finally, in Section \ref{sec:conclusions}, we summarize our results.


\section{Theoretical background}\label{sec:methods}

\subsection{Two-body problem in post-Newtonian theory}

When going to higher PN orders, calculations become notoriously algebraically intensive. It is therefore convenient to work in natural units, wherein $M = m_1 + m_2=1$, $G=1$, and $c=1$, where $m_1$ and $m_2$ are the masses of the two bodies, $G$ is the gravitational constant, and $c$ is the speed of light in vacuum. In these units, we can express all expansions of the PN series as Taylor expansions in the so-called PN parameter $x = -2E$, where $E$ is the Newtonian orbital energy per unit reduced mass $\mu = m_1m_2/M$. The PN parameter $x$ is also a gauge-invariant quantity \citep[see, e.g.][]{Blanchet2014}, meaning that the following calculations are valid regardless of gauge choice. With the units taken care of, two different ingredients are needed to fully describe the orbital evolution of a two-body system in PN theory.

Firstly, we require a parametrization of the orbital separation $r(\phi)$, which solves the PN equations of motion for a constant orbital energy $E$ and angular momentum vector $\mathbf{J}$:

\begin{align}
    r(\phi) &= r_{\rm N}(\phi) + r_{\rm{1\,PN}}(\phi) + r_{\rm{1.5\,PN}}(\phi) + r_{\rm{2\,PN}}(\phi) + ...,\\ \nonumber
\end{align}

\noindent where $\phi$ is the polar angle and the first term of the right-hand side is the solution to the familiar Kepler problem in Newtonian gravity. There are many equivalent possibilities to express such parametrizations, and in this paper we use the so-called quasi-Keplerian (QK) one \citep[see][]{1PN_par,2PN_par,Memmesheimer}, in which the radial separation is expressed in terms of  two generalised orbital parameters, a semi-major axis $a$ and an eccentricity $e$. These parameters depend on the quantities of the binary system, such as the energy $E$, the angular momentum vector $\mathbf{J}$, and, in the case of BHs, two spin vectors $\mathbf{S}_1$ and $\mathbf{S}_2$.\footnote{In this work, the magnitude of the spin vector of a BH of angular momentum's magnitude $L$ and mass $M$ is the dimensionless quantity defined by $0 \leq S \equiv c L /(G M^2) \leq 1$. Note that, if the secondary is not a BH, the definition of $\mathbf{S}_2$ needs to be revisited. For the scope of this paper, this clarification is not necessary, since we will neglect the spin of the secondary altogether.} They serve to greatly simplify the notation, just as introducing a semi-major axis and an eccentricity simplifies the solution of the Newtonian Kepler problem.

Secondly, we require a set of differential equations that describe how the aforementioned orbital parameters evolve due to the flux of energy and angular momentum caused by the emission of GWs. A convenient form is found when the orbital parameters $a$ and $e$ are expressed in terms of the PN parameter $x$ and a dummy eccentricity parameter $e_{\rm d}$, that essentially is a measure of the angular momentum of the system:

\begin{align}
a(x,e_{\rm d}) &= \frac{1}{x} \sum_{n=0}^{\infty} C_{n/2}^{a}(e_{\rm d}) x^{n/2}, \label{eq:a_series}\\
e(x,e_{\rm d}) &= \sum_{n=0}^{\infty} C_{n/2}^{e}(e_{\rm d}) x^{n/2}, \label{eq:e_series} \\ \nonumber
\end{align}

\noindent where the subscripts of the coefficients denote the PN order and some of the coefficients can vanish. The explicit coefficients used in this paper are listed in the Appendix, and can be found in \cite{Memmesheimer} for the spin-less part and in \cite{jetzer} for an extension that covers the spin-orbit coupling. The evolution equations for $x$ and $e_{\rm d}$ take a relatively simple form:

\begin{align}
\dot{x}(x,e_{\rm d}) &= x^5 \sum_{n=0}^{\infty} A_{n/2}(e_{\rm d}) x^{n/2}, \label{eq:xdot_series}\\
\dot{e}_{\rm d}(x,e_{\rm d}) &= x^4 \sum_{n=0}^{\infty} B_{n/2}(e_{\rm d}) x^{n/2},  \label{eq:eddot_series} \\ \nonumber
\end{align}

\noindent where, similarly to before, the subscripts of the coefficients denote the \textit{relative} PN order and some of the coefficients can vanish.\footnote{Note that (in this paper) the PN order of an energy or angular momentum flux is always relative to the quadrupole order contribution, which by itself is technically of 2.5~PN order.} The explicit coefficients used in this paper can be found in \cite{yannick} for the spin-less parts and in \cite{jetzer} for parts describing the spin-orbit coupling. Note that these evolution equations are always orbit-averaged, the assumption being that changes to the orbital elements occur on time-scales that are much longer than a single orbital period, which is the same assumption that is required to produce Peters' formula in the first place. A different approach would be to use a Hamiltonian integrator that evolves the positions of the binary's components step by step \citep[see, e.g.][]{2008AN....329..904B,2010PhRvD..81j4025L,2019PhRvD.100b4015I}. While a direct numerical integration of the equations of motion is in principle more precise, we chose to adopt the adiabatic approximation for two reasons. Firstly, we want to investigate a large range of parameter space in mass, mass ratio, orbital separation, eccentricity, and spin. The repeated integration of accurate equations of motion at high PN orders would simply be too costly. Secondly, IMRI/EMRI/XMRI sources complete thousands (if not millions) of orbits before coalescence, meaning that numerical errors, which accumulate orbit by orbit, would be very large. In those cases, the adiabatic approximation, whose PN terms are found through the Teukolsky formalism \citep{1973ApJ...185..635T,2007CQGra..24R.113A}, is more efficient and more accurate than orbital integrators, which are indeed often tested and calibrated on close to equal-mass ratio binaries. 

\subsection{Peters' time-scale and previous results}

At the lowest order in the PN parameter $x$, Equations~\eqref{eq:a_series} and \eqref{eq:e_series} reduce to $a= x^{-1}$ and $e = e_{ \rm d}$, respectively, which describe a binary moving along a Keplerian orbit. Equations~\eqref{eq:xdot_series} and \eqref{eq:eddot_series} reduce instead to $\dot{x}= A_0 x^5 f(e)$ and $\dot{e}= B_0 x^4 g(e)$, respectively, where $A_0$ and $B_0$ can be found in the appendix, and $f(e)$ and $g(e)$ are known as the eccentricity enhancement functions:

\begin{align}
    f(e)&=\left(1 + \frac{73}{24}e^2 +\frac{37}{96}e^4\right)(1-e^2)^{-7/2}, \\
    g(e)&=\left(1 + \frac{121}{304}e^2 \right)(1-e^2)^{-5/2}.
\end{align}

They describe a binary that is radiating GWs in accordance with \citet{GR}'s quadrupole formula. Because of the coupled nature of these equations and the complexity of the eccentricity enhancement functions, it is impossible to solve them analytically for an arbitrary initial semi-major axis $a_0$ and eccentricity $e_0$. Nevertheless, Peters and Mathews gave some approximated solutions for $a(t)$ and $e(t)$ \citep[see][]{Peters_Mathews_1963,Peters_1964}. In particular, the solutions for $a(t)$ can be set equal to zero and solved for the time $t$ in order to yield a decay time-scale of the orbital evolution.

The most commonly used formula for Peters' time-scale, $t_{\rm P}$, is derived by solving Equation~\eqref{eq:a_series} at lowest order while also neglecting the evolution of the eccentricity, thus setting $\dot{e}= 0$. The solution reads

\begin{align}
    t_{\rm P} = \frac{5(1 + q)^2}{256q} \frac{a_0^4}{f(e_0)}= \frac{5c^5(1 + q)^2}{256 G^3 M^3q} \frac{a_0^4}{f(e_0)}, \label{eq:pet}\\ \nonumber
\end{align}

\noindent where $q = m_1/m_2 \leq 1$ is the mass ratio, and we reintroduced the physical constants. As stated in the introduction, this formula is currently used in a wide variety of applications, even though it is known to fail for highly eccentric and for highly relativistic orbits.

In Paper~I, we found two simple factors, $R$ and $Q_{\rm f}$, that can be multiplied by Peters' time-scale to resolve its issues:

\begin{align}
t_{\rm P}(a_0,e_0) \to t_{\rm P}(a_0,e_0)R(e_0)Q_{\rm f}(a_0,e_0). \\ \nonumber
\end{align}

The fitting function $R(e_0) = 8^{ 1- \sqrt{1-e_0}}$ is purely Newtonian, and is needed to resolve the lack of eccentricity evolution in the standard formula. Other fits with the same purpose have been used in some select cases \citep[see, e.g.][]{Bonetti_et_al_2018}, but tend to be much more complex than the factor $R$ and are therefore uncommon.

The factor $Q_{\rm f}(a_0,e_0)$ is a simple fit that roughly accounts for both the more complex 1~PN parametrization given by Equations~\eqref{eq:a_series} and \eqref{eq:e_series}, and the relative 1~PN fluxes in Equations~\eqref{eq:xdot_series} and \eqref{eq:eddot_series}.
The corrected time-scale reads

\begin{align}
     t_{\rm P}RQ_{\rm f} = \label{eq:corrpet} 
     \underbrace{ \frac{5c^5(1 + q)^2}{256 G^3 M^3q} \frac{a_0^4}{f(e_0)} }_{\text{Peters' formula}}
    \underbrace{ 8^{ 1- \sqrt{1-e_0}} \exp \left( \frac{2.5 r_{\rm S}}{a_0(1-e_0)} \right)}_{\text{eccentricity and PN correction}},
\end{align}

\noindent where $r_{\rm S}$ is the effective \citet{Schwarzschild_1916} radius of the two-body system, $r_{\rm S} = 2GMc^{-2}$, and the coefficient 2.5 in the exponential function was chosen to fit a large range of initial eccentricities. In Paper~I, we showed that Equation~\eqref{eq:corrpet} can resolve errors of order 1--10 than can arise when applying Peters' formula to highly eccentric and relativistic orbits.

\subsection{Assumptions needed for higher orders}

BH spin couplings first appear at 1.5~PN order in the PN expansion. From then on, both spins have to be taken into account, drastically increasing the number of degrees of freedom and dynamical variables of the system. The spins can interact with the orbital angular momentum at 1.5 PN order and also with each other at 2 PN order. As is shown in Table~\ref{tab:table1}, this increases the total number of dynamical variables from 2 ($a$ and $e$) to 10. Sampling an appropriate range of parameter space with a grand total of ten initial conditions would simply be an exercise in frustration.

At the cost of some loss of generality, however, we can reduce the amount of new dynamical variables to only one. LISA BH sources will consist of three main categories: EMRIs, IMRIs, and SMBH binaries. These are mainly distinguished by ranges of their mass ratios: $\sim$10$^{-6}$, $\sim$$10^{-4}$ to $\sim$$10^{-2}$, and $\sim$10$^{-2}$ to $10^{0}$, respectively. For EMRIs and IMRIs, the mass ratio is by definition very small, but even for SMBH mergers only a small minority will have mass ratios of order one \citep[see, e.g.][where we assume that galaxy mass is a tracer for SMBH mass]{OLeary_et_al_2021}. In light of applications for LISA sources, it is very natural to restrict our analysis to small mass ratios, i.e. $q \lsim 0.1$. This turns out to be extremely convenient, because a small mass ratio suppresses most of the complex effects of spin-orbit and especially spin-spin couplings. We decided to also neglect the standard, non-Kerr 2~PN contributions to the parametrization and the fluxes of the two-body system. This last assumption is justified for two connected reasons: firstly, our numerical investigations showed that integer-PN orders tend to converge much faster than half-integer tail contributions \citep[see also][]{Blanchet2014}, making the 2~PN correction very small; secondly, our goal is to produce a correction to Peters' formula that is sufficiently simple to be used in the broader context of source modelling for GW observatories. Fitting the minute 2~PN deviations would necessarily require a much more complex fitting function.

\begin{table}
    \centering
    \begin{tabular}{c|c|c|c}
      Order & Parametrization  & Fluxes & Dyn. Variables \T \B \\
       \hline
        0~PN & Keplerian & Quadrupole & $a$, $e$ \T \B \\
        1~PN & 1~PN QK & 3.5~PN & $a$, $e$ \T \B \\
        1.5~PN & S-O  & S-O, PN tail & $a$, $e$, $\mathbf{\hat{J}}$,  $\mathbf{S}_1$, $\mathbf{S}_2$ \T \B \\
        2~PN & 2~PN, S-S  & 4.5~PN, S-S & $a$, $e$, $\mathbf{\hat{J}}$,  $\mathbf{S}_1$, $\mathbf{S}_2$ \T \B
    \end{tabular}
    \caption{This table summarises the new elements that enter the picture at a given PN order. The acronyms QK, S-O, and S-S stand for ``quasi-Keplerian'', ``spin-orbit'', and ``spin-spin'', respectively. After the first PN order, one must in principle consider the evolution of both BH spins, $\mathbf{S_1}$ and $\mathbf{S_2}$, as well as the specific angular momentum vector $\mathbf{\hat{J}}$.}
    \label{tab:table1}
\end{table}


\section{Vacuum time-scales}\label{sec:firstorder}

\subsection{Initial conditions and orbital evolution}

A divergence-free PN parametrization of the orbital motion that includes spin-orbit and spin-spin couplings can be found in \citet{jetzer}. Up to 1.5 PN order, it has the form of a standard QK parametrization:

\begin{align}
    r = a\left(1-e \cos{u} \right).
    \label{eq:spinparam}\\ \nonumber
\end{align}

The parameters $a$ and $e$ describe the shape of the orbit, whereas the parameter $u$ is responsible for orbital precession. They are expressed as series in the PN parameter $x$ and they read

\begin{align}
    a(x,e_{\rm d})&= \frac{1}{x} -\frac{7}{4} + 2\frac{\sqrt{x}}{\sqrt{1-e_{\rm d}^2}} \left\lvert \mathbf{\hat{J} \times S_1} \right\rvert + \mathcal{O}[x], \label{eq:ageneral}\\
    e(x,e_{\rm d})&= e_{\rm d} + 3 e_{\rm d} x - 2\frac{\sqrt{x^3}}{\sqrt{1-e_{\rm d}^2}} \left\lvert \mathbf{\hat{J} \times S_1} \right\rvert + \mathcal{O}[x^2], \label{eq:egeneral}\\ \nonumber
\end{align}

\noindent where the 1~PN term is taken from equations~(20a) and (20b) in \cite{Memmesheimer} and simplified according to the small mass-ratio limit. We wish to compare a Newtonian orbit and a PN orbit that  start out with the same orbital elements at a given moment in time ($t=0$) and are subsequently evolved with the appropriate energy and angular momentum fluxes. To enforce this, we set Equations~\eqref{eq:ageneral} and \eqref{eq:egeneral} equal to their Newtonian counterparts. The only way for these two orbits to be identical at $t=0$ is for them to have different values of the PN parameter $x$ and the dummy eccentricity $e_{\rm d}$. We denote them as ($x_{\rm N}$, $e_{\rm N}$) for the Newtonian orbit and as ($x_{\rm{PN}}$, $e_{\rm{PN}}$) for the PN orbit. To find the correct values for $x_{\rm{PN}}$ and $e_{\rm{PN}}$, we set the Newtonian and PN orbital elements equal to each other:

\begin{figure*}
    \centering
    \includegraphics[scale=0.48]{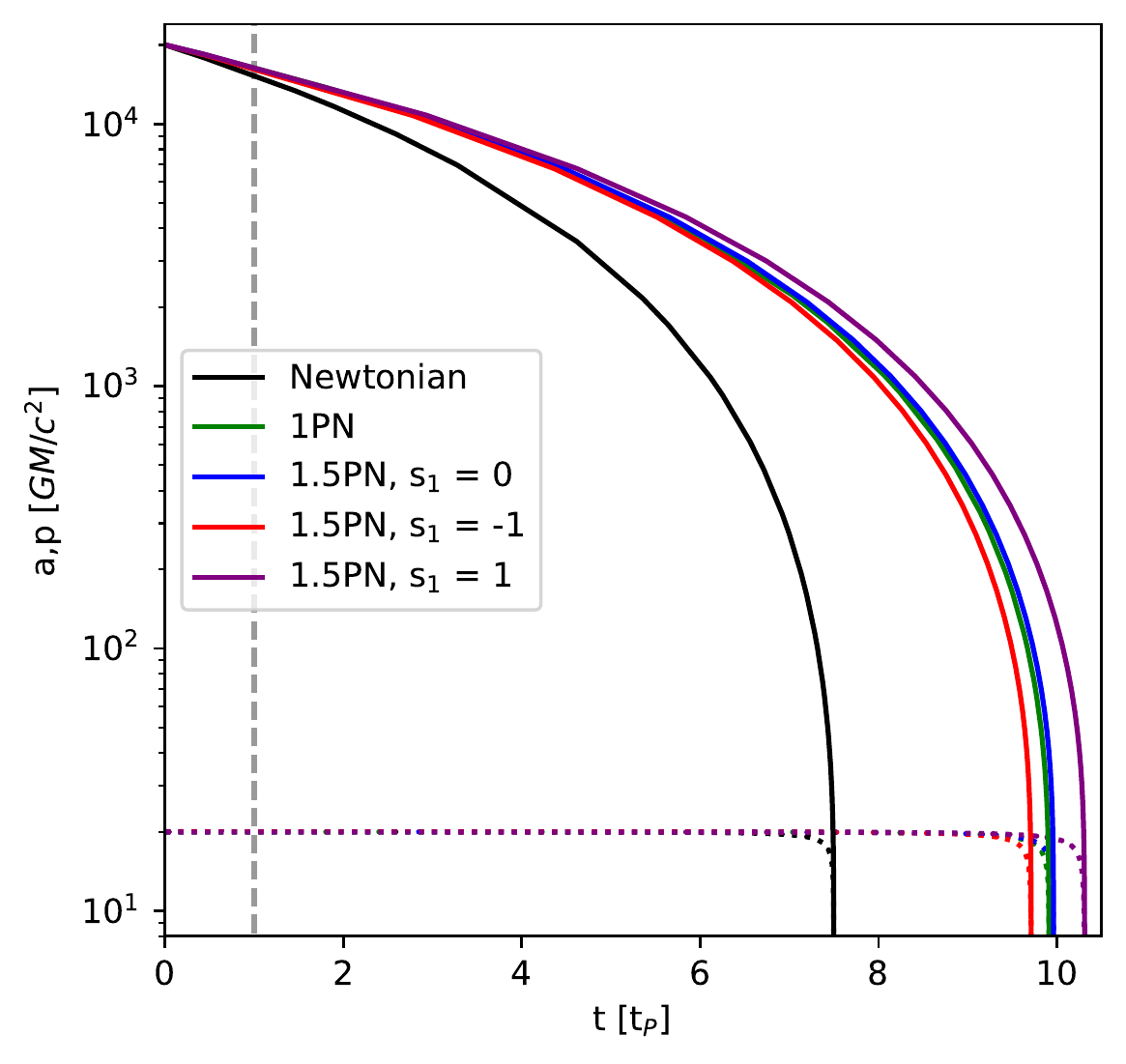}
    \includegraphics[scale=0.48]{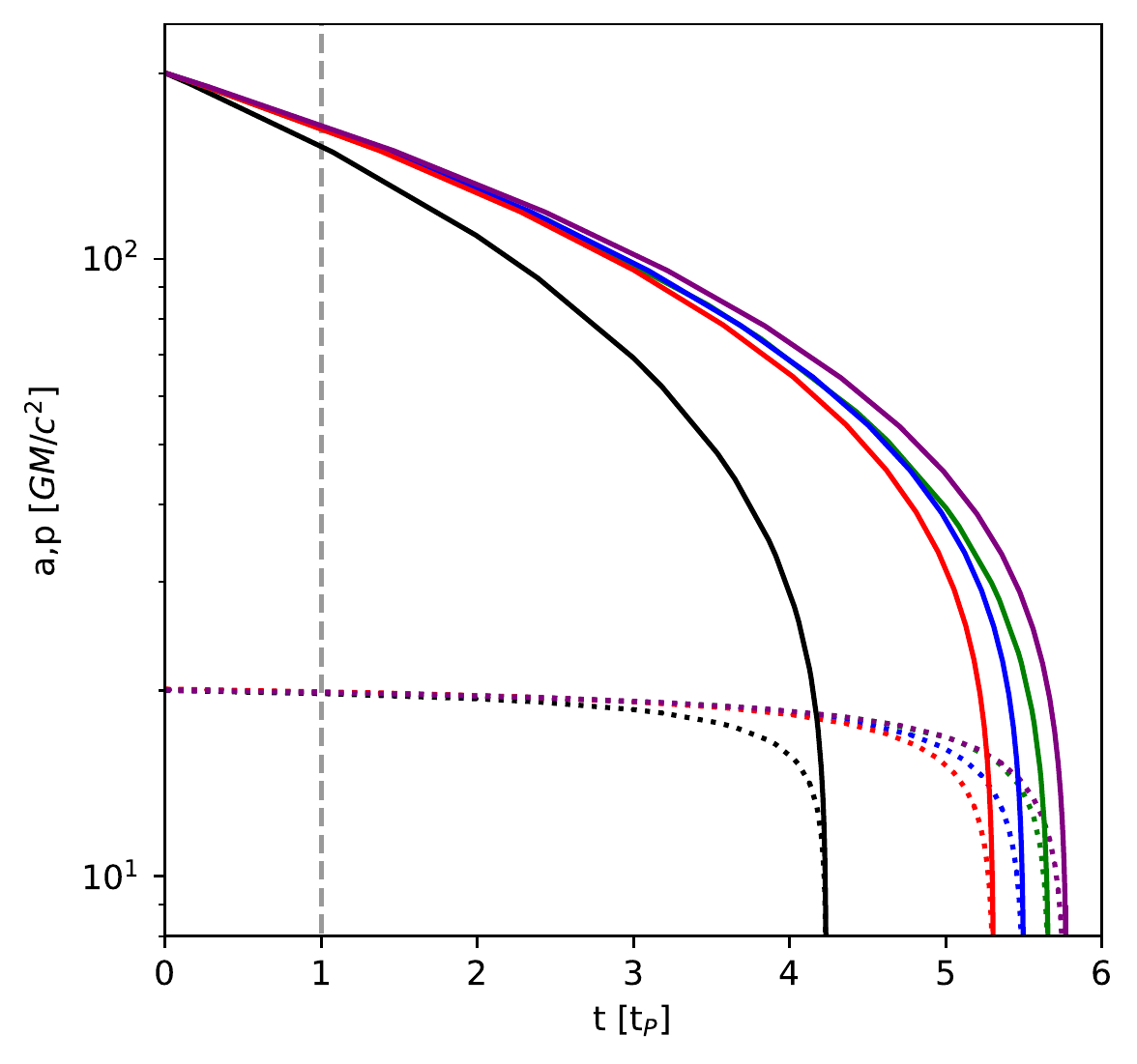}
    \includegraphics[scale=0.48]{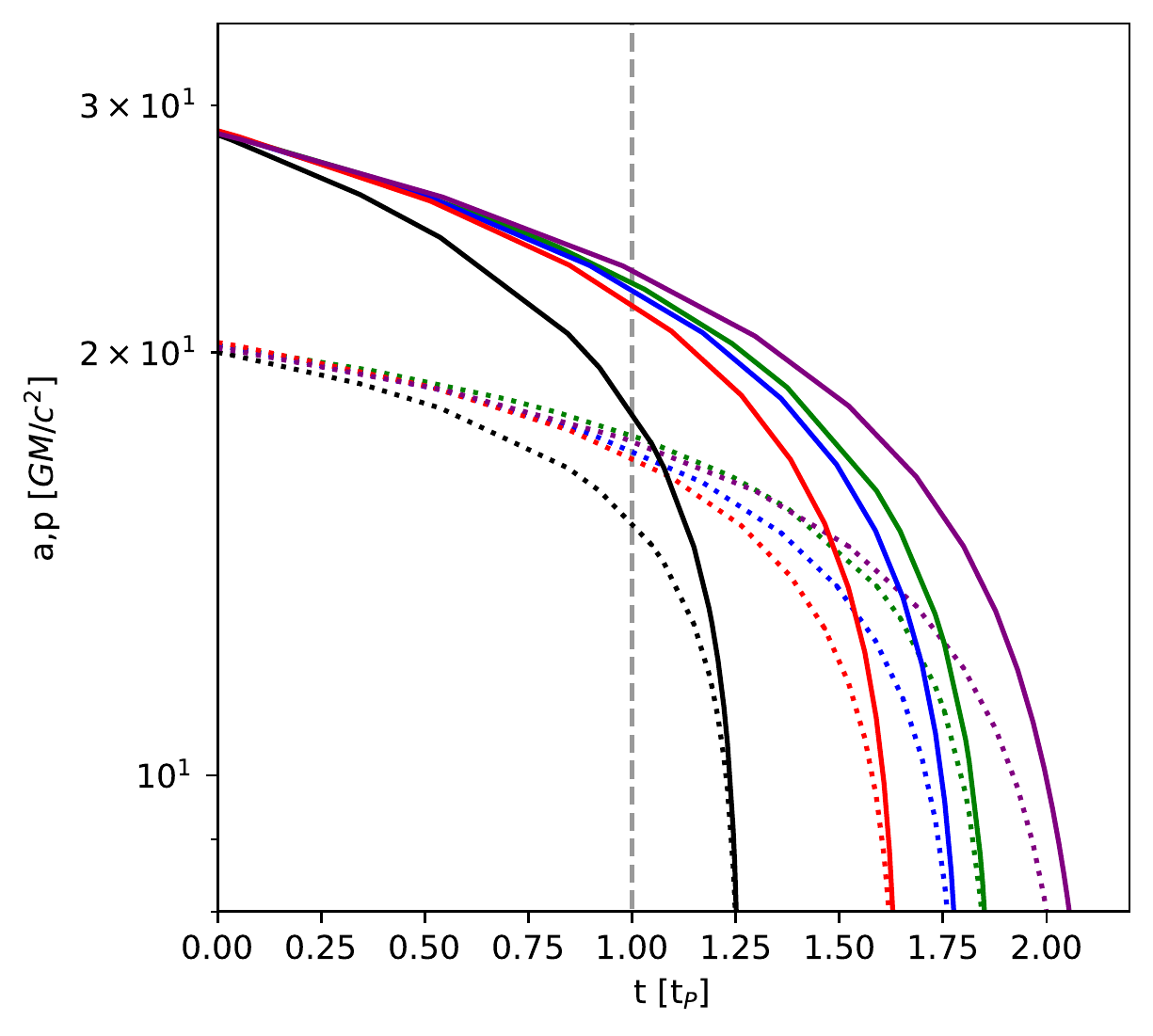}
    \caption{Three examples of the time evolution of the orbital semi-major axis $a$ (solid lines) and periapsis $p$ (dotted lines), for different initial eccentricities (0.999, 0.9, 0.3). The $x$-axis is scaled with Peters' time-scale, denoted by the vertical grey line. The different colours represent different PN orders and spin values for the primary BH. The spin values $s_1=-1$ (red) and $s_1=1$ (purple) are chosen to show the two extreme cases of an equatorial co-rotating and counter-rotating orbit around an extremal BH. The blue line ($s_1=0$) represents the Schwarzschild case with hereditary PN terms. Note how the higher-order PN corrections converge around the first order one much more strongly for eccentric orbits.}
    \label{fig:param}
\end{figure*}

\begin{align}
    \frac{1}{x_{\rm N}} &\overset{!}{=} a(x_{\rm{PN}},e_{\rm{PN}}), \\
    e_{\rm N} &\overset{!}{=} e(x_{\rm{PN}},e_{\rm{PN}}).\\ \nonumber
\end{align}

We can solve these equations order by order and find the values of the PN parameters so that the two orbits are identical at the beginning of their evolution:

\begin{align}
    x_{\rm{PN}} &= x_{\rm N} - \frac{7}{4}x_{\rm N}^2+ 2\frac{\sqrt{x_{\rm N}^5}}{\sqrt{1-e_{\rm N}^2}} \left\lvert \mathbf{\hat{J} \times S_1} \right\rvert + \mathcal{O}[x_{\rm N}^3], \label{eq:initialx} \\
    e_{\rm{PN}} &= e_{\rm N} - 3 e_{\rm N}x_{\rm N}+ 2\frac{\sqrt{x_{\rm N}^3}}{\sqrt{1-e_{\rm N}^2}} \left\lvert \mathbf{\hat{J} \times S_1} \right\rvert + \mathcal{O}[x_{\rm N}^2]. \label{eq:initiale}\\ \nonumber
\end{align}

As noted before, the energy and angular momentum fluxes can be averaged into secular evolution equations for the parameters $x$ and $e$. The equations are organised into PN series. Schematically, they read

\begin{align}
\dot{x} &= x^{5}\left(A_{\rm N} + xA_{\rm{1\,PN}} + x^{3/2} \left(A_{\rm{t}} + A_{\rm{SO}}\right) \right) + \mathcal{O}[x^7], \label{eq:dxdt} \\
\dot{e}_{\rm d} &= x^{4}\left(B_{\rm N} + xB_{\rm{1\,PN}} + x^{3/2} \left(B_{\rm{t}} + B_{\rm{SO}}\right) \right) +\mathcal{O}[x^6], \label{eq:dedt}\\ \nonumber
\end{align}

\noindent where the different subscripts denote the Newtonian and 1~PN contributions, as well as the lowest-order hereditary tail and the spin-orbit coupling contribution. The full expressions for the coefficients can be found in \cite{jetzer} and \cite{2019PhRvD.100h4043E} or in the Appendix.

In principle, we are missing various sets of evolution equations for the direction of the orbital angular momentum vector and the two spin vectors. However, at lowest order in the spin-orbit coupling and the mass ratio, the orbital angular momentum vector $\mathbf{J}$ only undergoes a Lense--Thirring \citep[][]{Lense_Thirring_1918} precession. Furthermore, all terms that contain $\mathbf{\hat{J}}$ are of the form

\begin{align}
\left\lvert \mathbf{\hat{J} \times S_1} \right\rvert= S_1 \cos{\theta} =s_1, \label{eq:effspin}\\ \nonumber
\end{align}

\noindent where $\theta$ is the angle between the primary BH spin vector and the orbital angular momentum vector. The magnitude of the BH spin is assumed to be constant, because the evolution occurs in vacuum. Furthermore, the Lense--Thirring precession does not affect the angle $\theta$, hence the expression in Equation~\eqref{eq:effspin} is constant in time. It follows then that Equations~\eqref{eq:dxdt} and \eqref{eq:dedt} are a closed system of differential equations, which can be integrated numerically, starting from the correct initial conditions given by Equations~\eqref{eq:initialx} and \eqref{eq:initiale}.

\begin{figure*}
    \centering
    \includegraphics[scale=0.7]{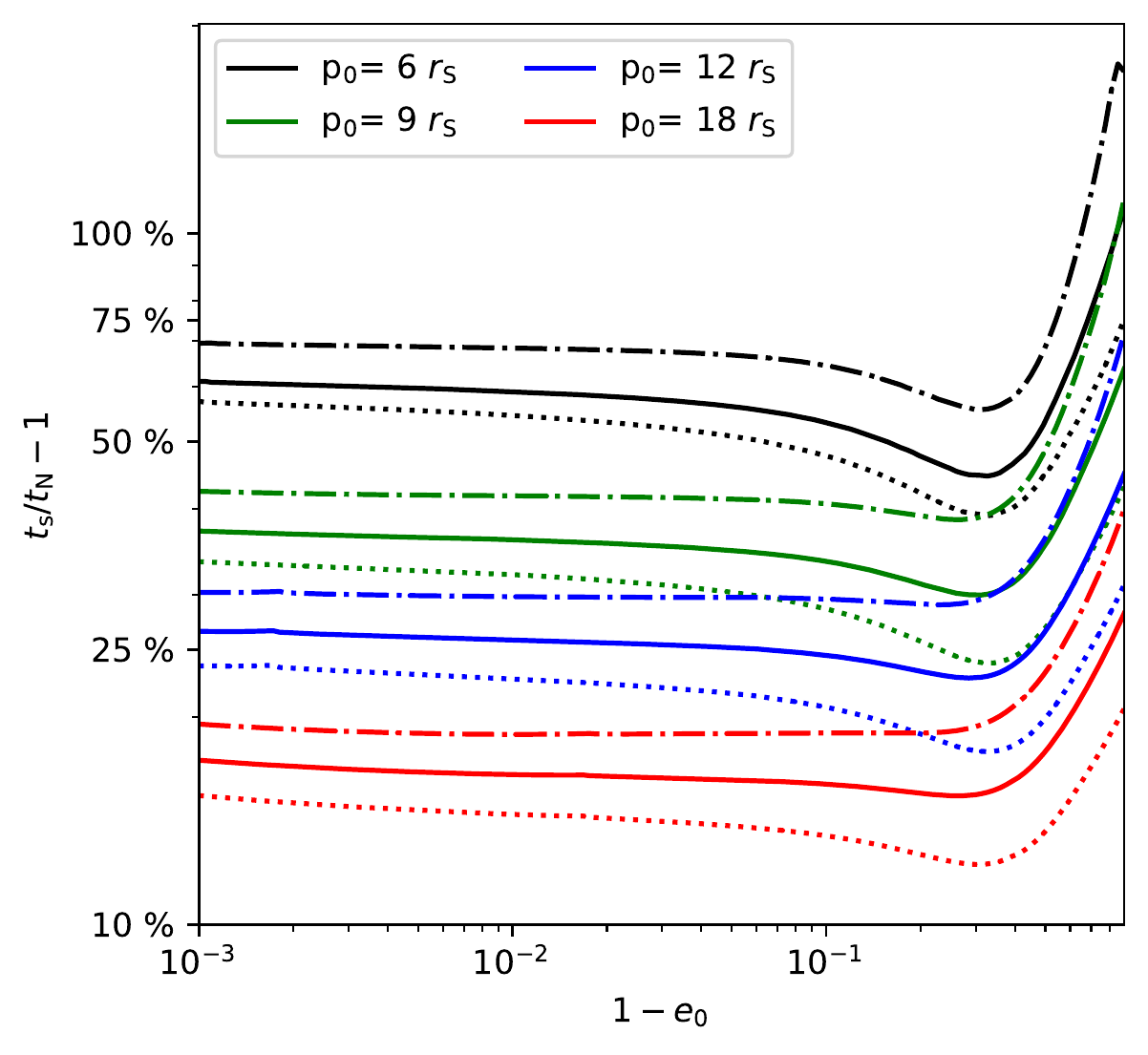}
    \includegraphics[scale=0.7]{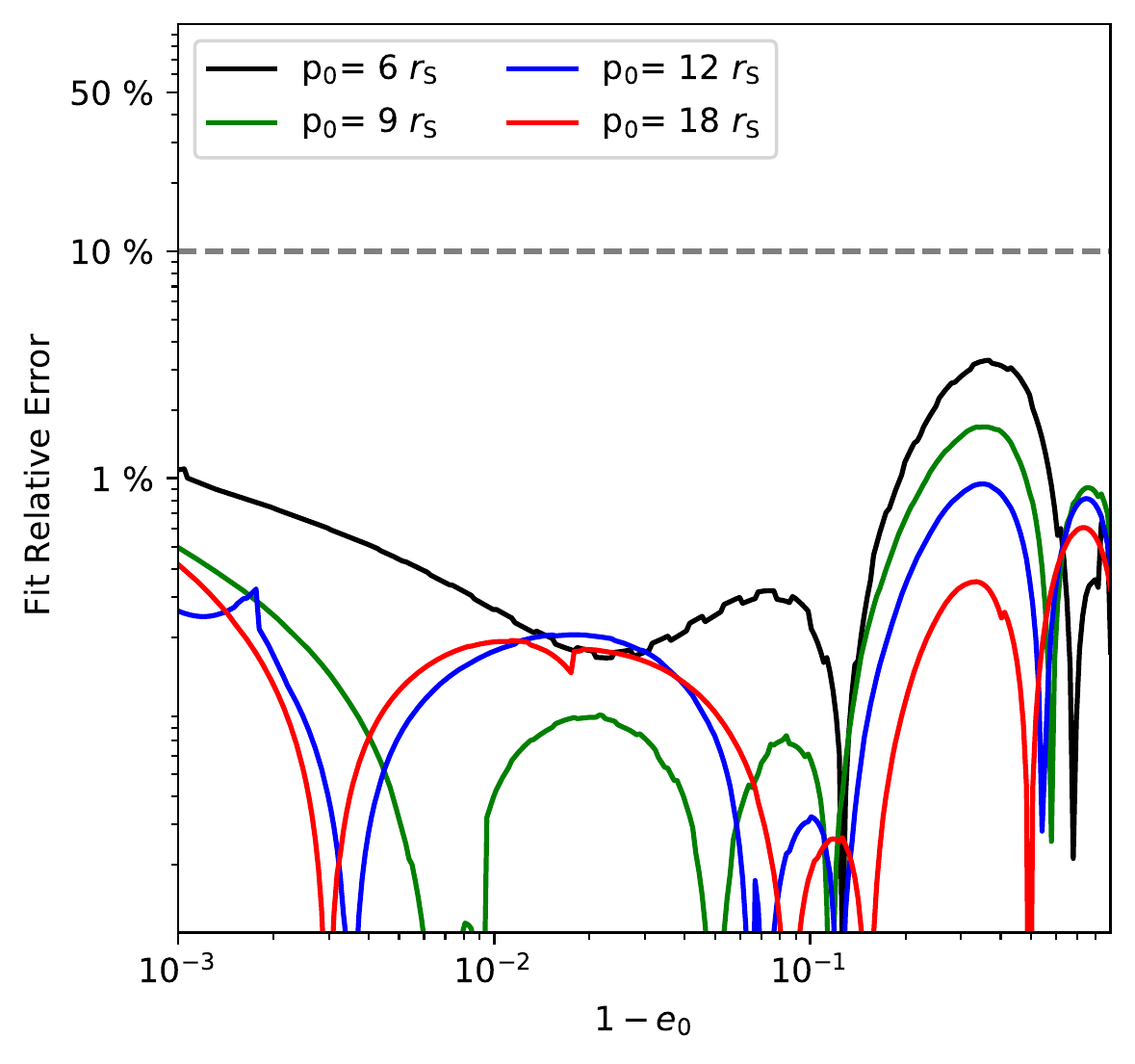}
     \caption{In the left-hand panel, we plot the relative increase of the time-scale of GW-induced decay due to the 1~PN and 1.5~PN contributions, $t_{\rm s}$, with respect to the Newtonian prediction, $t_{\rm N}$, for different values of the initial periapsis. The solid lines represent the spinless case, whereas the dotted and dash-dotted lines represent the minimum ($-1$) and maximum (1) values for the spin parameter $s_1$, respectively. In the right-hand panel, we show the relative error of the fits we propose for the same range of parameter space. For the sake of clarity, we avoid plotting the curves with spin, but note that the fits perform similarly to the spinless case (with the single exception of a circular, equatorial and prograde orbit at $6 r_{\rm S}$ around a maximally spinning primary BH, where the fit fails by no more than 30 per cent). For a periapsis $p_0 \gsim 10 r_{\rm S}$, the errors lie well below 1 per cent. Even for the extreme case of $p_0 = 6 r_{\rm S}$ the fits perform rather well, considering the fact that the PN series itself is known to break down below those separations.}
    \label{fig:ratioPN}
\end{figure*}

As we have noted before, there are two separate effects that influence the inspiral time-scale when considering higher-order PN perturbations. On the one hand, the evolution equations for the orbital parameters gain additional terms, which originate from the additional energy and angular momentum fluxes and, generally, tend to shorten the inspiral time-scale. On the other hand, some adjustments to the initial conditions are required to ensure that the orbital elements of a PN orbit and those of a Newtonian orbit are identical at the beginning of the evolution, which tend to increase the overall time-scale. The reason for this increase is that a binary in a PN gravity field requires slightly more energy and angular momentum to support a given orbital separation. This extra energy must be radiated away and therefore tends to increase the overall inspiral time-scale. To get a feel for how different orders affect the overall behaviour, we show the evolution of a highly eccentric ($e_0=0.999$), moderately eccentric ($e_0=0.9$), and low-eccentricity ($e_0=0.3$) orbit in Figure~\ref{fig:param}. In all cases, Peters' formula underestimates the inspiral time-scale, especially so for eccentric orbits. The effect of the 1 PN correction is also to further increase the time-scale, whereas higher-order corrections cluster around the 1 PN result.

\subsection{Correction ratios}\label{sec:CorrectionRatios}

In order to compute the inspiral time-scale, we performed numerous numerical integrations of the orbital evolution, sampling a wide range of initial orbital separations, eccentricities, and spin parameter values. We define the merger event as the moment at which the periapsis of the orbit, $p = a(1-e)$, reaches the effective Schwarzschild radius. We express the corrections to Peters' time-scale as ratios $R$ and $Q_{i}$, where the index $i$ stands for the several different effects that we want to describe. The most complete and accurate formula for the time-scale is obtained by multiplying Peters' formula $t_{\rm P}$ by three separate correction ratios,

\begin{align}
    t_{\rm P}(a_0,e_0) \to t_{\rm P}(a_0,e_0)R(e_0) Q_{\rm h}(a_0,e_0) Q_{\rm s}(a_0,e_0,s_1), \\ \nonumber
\end{align}

\noindent where the ratio $R(e_0)$ accounts for the secular eccentricity evolution at quadrupole order, $Q_{\rm h}(a_0,e_0)$ accounts for the 1~PN and 1.5~PN spinless effects, and $Q_{\rm s}(a_0,e_0,s_1)$ accounts for the inclusion of the spin-orbit coupling. We calculate these ratios numerically by integrating the appropriate evolution equations, and subsequently find various fits for the results. The numerical results can be seen in the left-hand panel of Figure \ref{fig:ratioPN}, where we plot the ratio of the 1.5 PN time-scale to the Newtonian time-scale for different values of the initial periapsis, spin, and initial eccentricity. As shown in Paper I, the correction generally grows for smaller periapsis, while the two edge cases of co- and counter-rotating orbits around an extremal BH envelop the spin-less results. In the circular case, the spin parameter $s_1$ is degenerate with the periapsis of the orbit by $\gsim 5 r_{\rm S}$, while it has much less of an influence in the eccentric case. Our proposal for the appropriate fits to these ratios reads

\begin{align}
R(e_0) &= 8^{1- \sqrt{1-e_0}}, \label{eq:r1} \\
Q_{\rm h}(p_0,e_0) &= \exp \left(\frac{2.8 r_{\rm S}}{p_0} \right) f(p_0,e_0), \\
f(p_0,e_0)&=\left( 1 +  A(p_0)\left( 1 - e_0 \right)^2 + B(p_0)\left( 1 - e_0 \right) \right), \nonumber \\
A(p_0)&= \left( -1 + \exp \left(\frac{2.2 r_{\rm S}}{p_0} \right) + \left( \frac{3.8 r_{\rm S}}{p_0} \right)^{3/2} \right), \nonumber \\
B(p_0)&= - \left( \frac{3.8 r_{\rm S}}{p_0} \right)^{3/2}, \nonumber \\
Q_{\rm s}(p_0,e_0,s_1) &= \exp \left( A_{\rm s}(p_0,e_0) s_1  + B_{\rm s}(p_0,e_0)\lvert s_1 \rvert ^{3/2} \right), \label{eq:r2}\\
A_{\rm s}(p_0,e_0) &= e_0 \frac{0.3 r_{\rm S}}{p_0} + (1-e_0)^{3/2} \left( \frac{3.7 r_{\rm S}}{p_0}\right) ^{3/2}, \nonumber \\
B_{\rm s}(p_0,e_0) &= \left( e_0 \frac{1.1 r_{\rm S}}{p_0} \right)^{5/2} + (1-e_0)^{3} \left( \frac{4.3 r_{\rm S}}{p_0}\right) ^{3}, \nonumber \\ \nonumber
\end{align}

\noindent valid for arbitrary values of the initial semi-major axis, eccentricity, and (primary BH) spin parameter. As shown in Figure \ref{fig:ratioPN}, we find these fits to be an accurate estimate of the numerical results for orbits whose initial periapsis is no lower than $p_{0} \sim 3r_{\rm s} (2 -e_0)$. Below that point, the assumptions of the PN expansion break down, and we cannot expect our results to be accurate.

These general correction factors turn out to be rather complicated and unwieldy. The reason for this is the complicated behaviour of the time-scale in a region of parameter space between $0.1 \lsim e_0 \lsim 0.8$, visible in Figure~\ref{fig:ratioPN}. The likely cause for this are some terms in the higher-PN fluxes switching sign at eccentricities of $\sim$0.5, complicating the dynamics for orbits that start evolving from that region of phase space. With regards to LISA sources, however, most formation channels do not predict an isotropic distribution of eccentricities, but rather a strong preference for either highly eccentric or almost circular orbits. Let us take EMRIs as an example: on the one hand, dynamical processes tend to generate EMRIs with $e_0 \gg 0.9$ by scattering compact objects onto almost radial orbits \citep[see, e.g.][]{Amaro-Seoane2007,pau_2013}; on the other hand, some EMRIs can be produced when a binary containing a compact object is disrupted by the tidal influence of the SMBH \citep[see, e.g.][]{Amaro_Seoane_2019}. These would likely enter the GW-dominated phase with very low eccentricities. For this reason, the most general form for the fits of the correction ratios is often not required. Rather, one can freely choose an appropriate eccentricity range or limit that is expected for a particular formation channel. Here we will discuss the two most obvious limits of $e_0 \to 0$ and $e_0 \to 1$. In these limits, the fits reduce to much more manageable formulas that satisfy our goal to be as simple as possible.

\subsubsection{Quasi-circular orbits}

If the binary enters the GW-dominated phase with a very low eccentricity ($e_0 \lsim 0.1 $), the fits for the correction ratios simplify to the following formulas:

\begin{align}
    R &\to 1, \\
    Q_{\rm h}Q_{\rm s} &\to \exp \left( \frac{5.0 r_{\rm S}}{p_0} + s_1\left( \frac{3.7 r_{\rm S}}{p_0}\right) ^{3/2} + \lvert s_1 \rvert ^{3/2}\left( \frac{4.3 r_{\rm S}}{p_0}\right) ^{3} \right). \label{eq:r3} \\ \nonumber
\end{align}

In this case, the eccentricity correction factor does not have a significant effect, but the PN correction factors take their maximal values for the coefficients. This occurs because binaries on circular orbits spend the whole duration of the orbit at small separations, maximising the influence of higher-order PN effects. In this limit, the fits are accurate to $\sim$10 per cent of the relative value if $p_0 \gsim 6 r_{\rm S}$ (see Figure~\ref{fig:ratioPN}).

\subsubsection{Highly eccentric orbits}\label{sec:HighlyEccentricOrbits}

If the binary enters the GW dominated phase with a very high eccentricity ($e_0 \gsim 0.8 $), the fits for the correction ratios simplify to the following formulas:

\begin{align}
 R &= 8^{1-\sqrt{1-e_0}}, \\
    Q_{\rm h}Q_{\rm s} &\to \exp \left( \frac{2.8 r_{\rm S}}{p_0} + s_1\left( \frac{0.3 r_{\rm S}}{p_0}\right) + \lvert s_1 \rvert ^{3/2} \left( \frac{1.1 r_{\rm S}}{p_0} \right)^{5/2} \right). \label{eq:r4} \\ \nonumber
\end{align}

In this case, the eccentricity correction factor dominates, whereas the PN correction factors are slightly suppressed with respect to the circular case. Nonetheless, both factors can compound on each other to reach values of $\sim$10 for  large regions of initial parameter space. In this limit, the fits are accurate to $\sim$10 per cent of the relative value if $p_0 \gsim 3 r_{\rm S}$, and to $\sim$1 per cent if $p_0 \gsim 6 r_{\rm S}$ (see Figure~\ref{fig:ratioPN}).

\subsection{Application: EMRI critical semi-major axis}

\begin{figure}
    \includegraphics[scale=0.7]{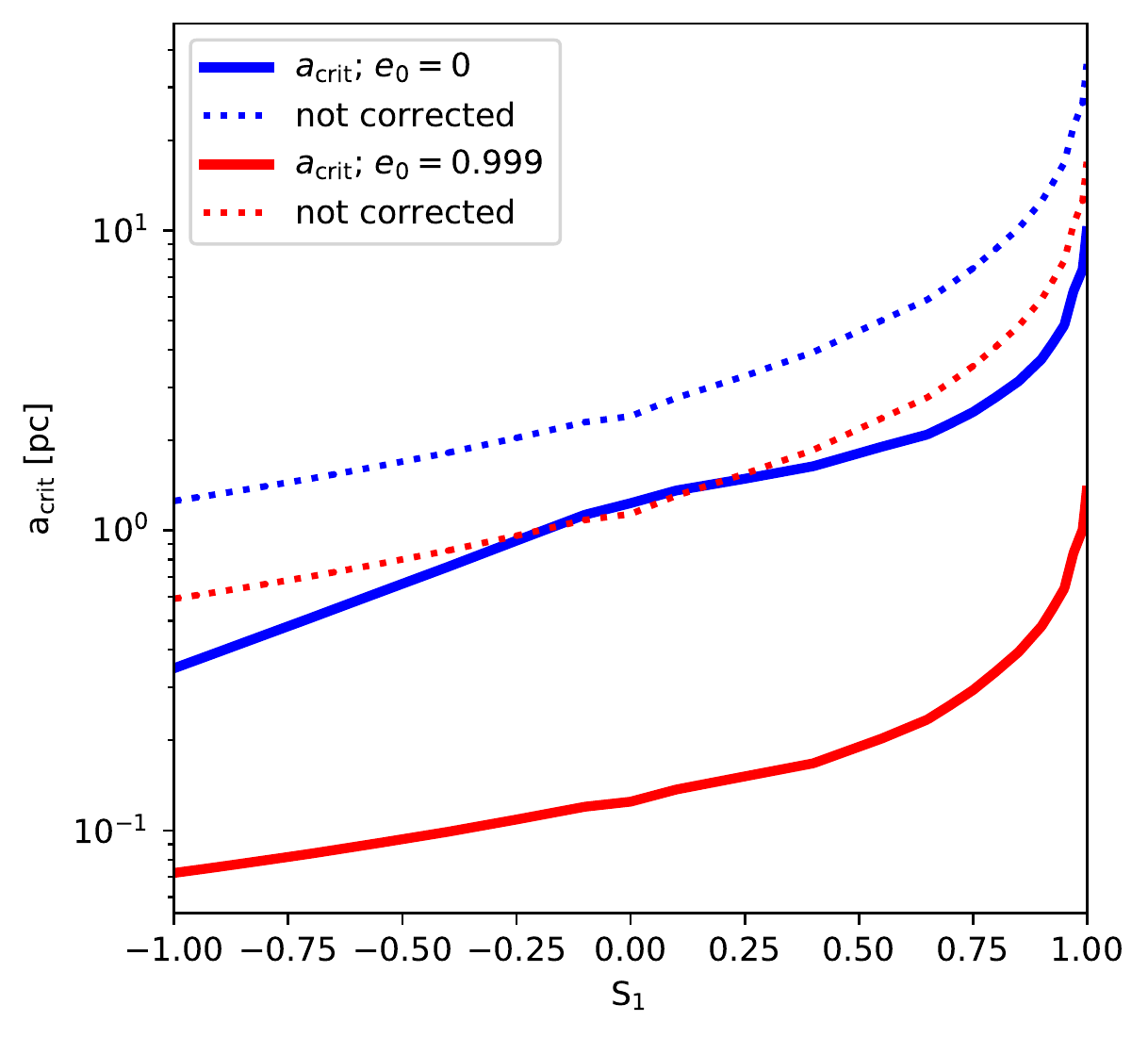}
    \caption{We plot the corrected critical semi-major axis $a_{\rm{crit}}$ (solid lines) as a function of the SMBH spin for an EMRI composed of a $4 \times 10^6$~M$_{\sun}$ SMBH and a BH of 10~M$_{\sun}$, for two initial eccentricities. The results are shown for orbits with a nominal inclination of $\theta=0.4$ with respect to the SMBH equatorial plane, chosen to represent a typical case. The effect of the correction factors to Peters' time-scale is to shift the curves downwards, and also to change the scaling with the SMBH spin.} 
    \label{fig:acritVSspinHIGHecc}
\end{figure}

The EMRI critical semi-major axis, $a_{\rm crit}$, marks the separation between the dynamical-evolution regime and the GW-emission regime, and is an essential ingredient for the calculation of EMRI event rates. The value of $a_{\rm crit}$ is found by equating the inspiral time-scale to the dynamical relaxation time-scale for a periapsis equal to the primary's last stable parabolic orbit. In Figure~\ref{fig:acritVSspinHIGHecc}, we plot $a_{\rm crit}$ as derived in \cite{Amaro_Seoane_2019}, this time including the correction factors to the GW-inspiral time-scale, for a system of a 10 M$_{\sun}$ BH inspiralling into a $4 \times 10^6$~M$_{\sun}$ SMBH.

For eccentric orbits, we find that the value of the critical semi-major axis decreases by an order of magnitude from previous results that use Peters' formula as a simple model of the inspiral time-scale. In the circular orbit case, the value of $a_{\rm crit}$ is generally decreased by a factor of a few, but the effects of the spin correction factor $Q_{\rm s}$ are clearly visible in the steeper scaling of the counter-rotating case ($S_1 < 0$), and in the slightly flattened scaling for the co-rotating case ($S_1 > 0$). The difference between the circular and the eccentric cases is also increased by almost an order of magnitude, which suggests different relative efficiencies for formation channels that preferentially create circular versus eccentric EMRIs. Note that the ISCO of a BH in a prograde orbit can be as low as $\gsim$0.5$ r_{\rm S}$, which is lower than the value where our fits (and PN theory in general) are accurate. In these cases, we evaluate the correction factors at $3 r_{\rm S}$ for highly eccentric orbits and at $6 r_{\rm S}$ for circular orbits, to ensure that the results do not diverge. This would represent a conservative estimate of the real change of the critical semi-major axis.

The expected event rates for EMRIs are then computed by performing an integral over phase space, in which $a_{\rm{crit}}$ is one upper limit \citep{pau_2013,Amaro_Seoane_2019}. The total phase-space volume of integration scales as $a_{\rm{crit}}^3$, which means that even moderate changes to $a_{\rm{crit}}$ can potentially have a large effect on the expected event rates. For this reason, we are currently working on a precise re-derivation of EMRI and XMRI \citep{Amaro_Seoane_2019} event rates that include the corrections to the inspiral time-scale. Preliminary results show that the rates decrease by more than one order of magnitude when switching from Peters' time-scale to the corrected formula. However, note that some event-rate estimates compute the critical semi-major axis using a version of the merger time-scale that focuses specifically on highly eccentric orbits \citep{Amaro_Seoane_2019}. In that case, our correction to the critical semi-major axis is only due to PN effects and the rates generally vary by a factor of a few.

\section{Vacuum time-scales versus environmental perturbations} \label{sec:environment}

\subsection{Perturbations by a third body}

In this section, we consider the case in which a distant massive body perturbs the orbit of an inspiralling LISA source. In other words, we consider a hierarchical three-body problem in which the hard binary is made of compact objects at very small orbital separations, and can exhibit relativistic behaviour. Following \cite{Will_KL}, we express the energy of such a system in the following way:

\begin{align}
    E = E_{\rm N} + E_{\rm{P}} + E_{\rm{1\,PN}} + \mathcal{O}\left[\frac{m_3}{c^2}\right],\\ \nonumber
\end{align}

\noindent where $E_{\rm N}$ and $E_{\rm{1\,PN}}$ are the Newtonian and 1 PN energy contributions, whereas $E_{\rm{P}}$ is the change in energy of the hard binary of a given osculating semi-major axis and eccentricity due to the influence of the distant third body. The explicit expression for $E_{\rm{P}}$, in the case of a perturber in a circular orbit, reads

\begin{align}
   E_{\rm{P}} &= -\frac{1}{4} \frac{G \mu m_{ 3} p_{0}^2}{R^3}\left(1 -3 \sin^2 \left( \mathcal{I}_0 \right) \sin^2 \left( \omega_0 \right) \right) \nonumber \\
   &= -\frac{1}{4} \frac{G \mu m_{ 3} p_{0}^2}{R^3} \mathcal{G}(\mathcal{I}_0,\omega_0) \\ \nonumber
\end{align}

\noindent where the function $\mathcal{G}$ describes the angle dependence of the energy shift, $\mu = m_1m_2/M$, $m_3$ is the mass of the perturber, $R$ its distance to the centre of mass of the hard binary, $\mathcal{I}_0$ the inclination with respect to the orbital plane of the perturber, and $\omega_0$ the angle of the periapsis with respect to the ascending or descending nodes of the orbital plane of the perturber. The remarkable fact proven in \cite{Will_KL} is that this expression remains constant over time-scales at which perihelion precession can significantly change the orientation of the orbit with respect to the plane of the perturber. The terms that would normally vary in $\omega$ are cancelled by cross terms between the PN dynamics of the inner binary and the tidal forces induced by the perturber. Indeed, the second $\sin^2$ term is evaluated at an \textit{initial} perihelion orientation, and therefore is not affected by perihelion precession. This result is very convenient for our purposes, because it is essentially of the same form as the change in initial conditions caused by a PN perturbation.

We can estimate how the change in energy of the hard binary caused by the perturber will affect the inspiral time-scale by readjusting the initial conditions of the GW-induced decay. We define a fictitious semi-major axis $a_{\rm {P}}$ and eccentricity $e_{\rm{P}}$, constructed from both the Newtonian energy of the hard binary, $E_{\rm N}$, and the constant shift induced by the perturber, $E_{\rm P}$:

\begin{align}
    a_{\rm P} &= -\frac{G M}{2\left(E_{\rm N} +  E_{\rm {P}}\right)} \nonumber \\ &\approx a - \frac{m_3 \mu a^4}{2 M R^3}(1-e)^2\mathcal{G}(\mathcal{I}_0,\omega_0),\\
    e_{\rm{P}}^2 & \approx e^2 - \frac{E_{\rm{P}}}{E_{\rm{N}}}(1-e^2),\\ \nonumber
\end{align}

An estimate for the change in the inspiral time-scale can by found by evaluating a particular vacuum time-scale $t_{\rm{G W}}$ at the fictitious semi-major axis and dividing it by the corresponding vacuum time-scale of the physical orbit:

\begin{align}
\frac{t_{\rm{G W}}(a_{\rm{P}},e_{\rm{P}})}{t_{\rm{G W}}(a,e)} &\approx \left(\frac{E_{\rm N} +  E_{\rm {P}}}{E_{\rm N}} \right)^{-4} \approx 1 - 4 \frac{E_{\rm {P}}}{E_{\rm N}}\mathcal{F}(e) \nonumber
\\ &\approx 1 - 2\frac{m_3 p^3 \mathcal{G}(\mathcal{I}_0,\omega_0)}{ M R^3 (1-e)}\mathcal{F}(e)\\ \nonumber
\end{align}

\noindent where, in the first step, we assumed quadrupole radiation or, equivalently, the evolution equations of \cite{Peters_Mathews_1963}. The function $\mathcal{F}$ can be computed from the eccentricity dependence of the inspiral time-scale, and ranges from 1/5 to 1/2 for most values of the eccentricity. From the results of Paper~I and previous sections, we know that PN corrections to the Newtonian vacuum time-scale can be described by factors of the form $\exp \left( C_{\rm PN} r_{\rm S}p_0^{-1} \right) \approx 1 + C_{\rm PN} r_{\rm S}p_0^{-1} + \mathcal{O}[(r_{\rm S}p_0^{-1})^n]$, where $C_{\rm PN}$ is a coefficient of order unity and $n$ is the PN order. We can therefore estimate the radius at which the environmental perturbation becomes of Newtonian (or PN) order by solving the equation

\begin{align}
    \frac{2 m_3 p_{0}^3 \mathcal{G}(\mathcal{I}_0,\omega_0)}{ M R^3 (1-e)}\mathcal{F}(e) = C_{\rm{PN}} \left( \frac{r_{\rm S}}{p_0} \right)^{n}\\ \nonumber
\end{align}

\noindent and find the characteristic radii $R_{\rm{P/PN}}$ at which the effect of the perturbation becomes as large as that of a given PN order:

\begin{align}
    R_{\rm{P/PN}} =\left( \frac{2\mathcal{G}(\mathcal{I}_0,\omega_0)m_3p_0^{3+n} }{(1-e)Mr_{\rm S}^{n}}\right)^{1/3}, \label{eq:rpn}\\ \nonumber
\end{align}

\noindent where we approximated $\mathcal{F}(e)^{1/3} \approx 1$ and $C_{\rm{PN}}^{1/3} \approx 1$. For the Newtonian order ($n=0$ and $C_{\rm{ PN}}=1$), the radius reads

\begin{align}
    R_{P/\rm N}\approx \mathcal{G}(\mathcal{I}_0,\omega_0) ^{1/3}\left(\frac{m_3}{M} \right)^{1/3}a_0(1-e_0)^{2/3},\\ \nonumber
\end{align}

\noindent which can only be of order of or smaller than the semi-major axis of the hard binary itself, breaking the assumption of a hierarchical triplet. In general, every additional PN order only adds a factor of a few over the previous value and, since we are considering hierarchical triplets (where $R \gg a_0$), we can conclude that the perturbation has very little influence on the inspiral time-scale. 

\begin{figure}
    \centering
    \includegraphics[scale=0.7]{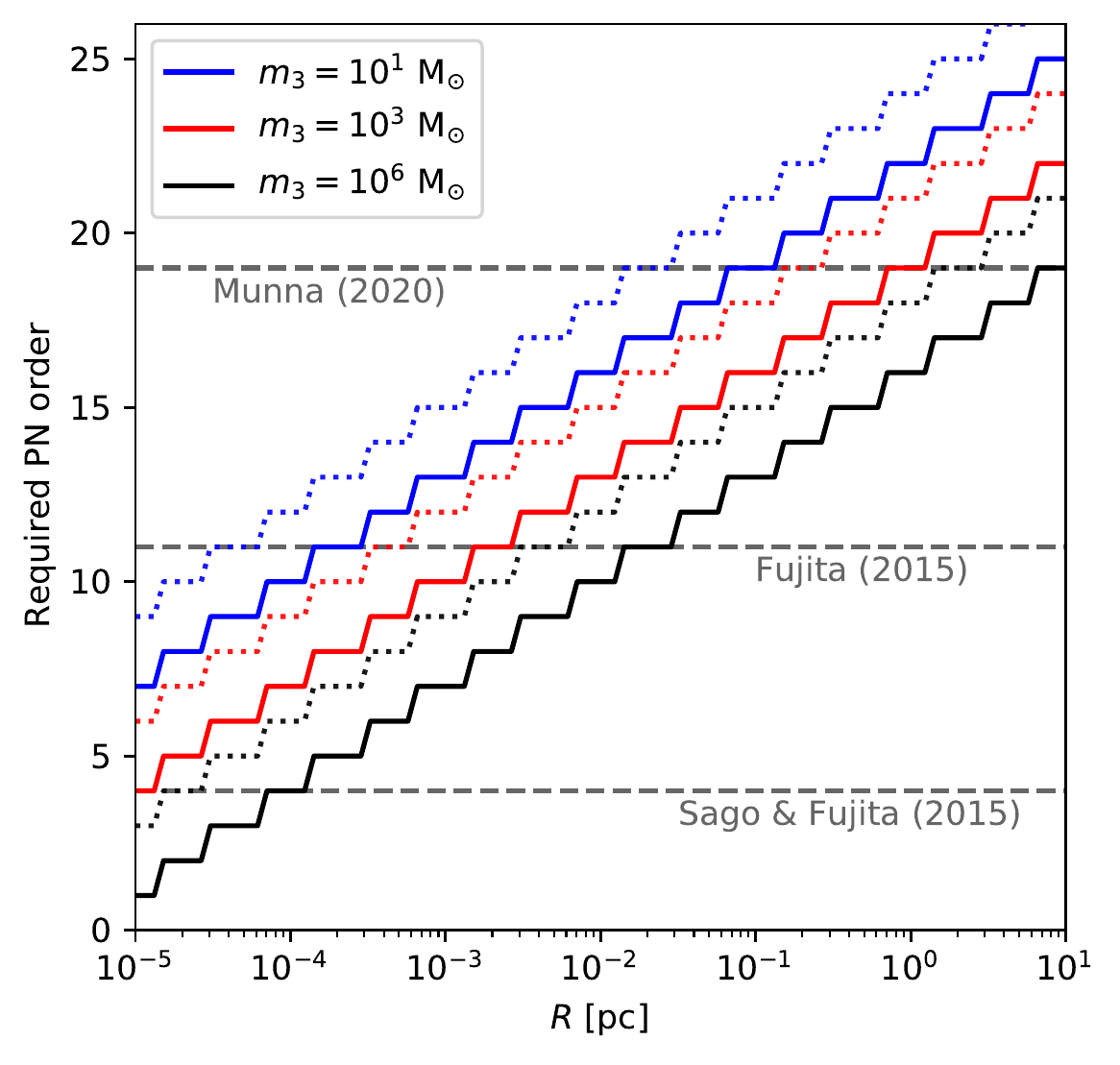}
    \caption{We show the minimum required PN order of a waveform template needed to detect a shift in the inspiral time-scale of an EMRI with an orbital frequency in the LISA band ($f\approx 1 $ mHz) caused by the perturbation of a third body with mass $m_3$, in the case of an inner binary composed of an SMBH of $10^6$~M$_{\sun}$ and a compact object of 10~M$_{\sun}$. We calculate the value by solving Equation~\eqref{eq:rpn} for the PN order $n$ and adding one. The solid lines represent a quasi-circular inspiral, whereas the dotted lines represent orbits that have the same frequency but an eccentricity of 0.9. The horizontal lines stand for the state-of-the-art results in the calculations of PN fluxes using different methods, for spinning BHs in eccentric (4~PN) and circular (11~PN) orbits \citep{2015PTEP.2015c3E01F,2015PTEP.2015g3E03S}, and also eccentric orbits for non-spinning BHs \citep{2020arXiv200810622M}.}
    \label{fig:pntempl}
\end{figure}

On the other hand, if the accumulated signal-to-noise ratio (SNR) of the source is high enough, LISA might be able to detect deviations from very high PN orders. In Figure~\ref{fig:pntempl}, we show the minimum order of a PN-waveform template that is required to detect a shift in the time-scale for a perturber of a given mass and separation, in the case of an inner binary composed of an SMBH of $10^6$~M$_{\sun}$ and a compact object of 10~M$_{\sun}$. We compute it by solving Equation~\eqref{eq:rpn} for the PN order $n$ and adding one. If we assume that the SNR is sufficient to reach current state-of-the-art waveform templates with spin \citep[depicted in Figure~\ref{fig:pntempl};][]{2015PTEP.2015c3E01F}, we find values that are similar to the results in \cite{PhysRevD.83.044030}, wherein it is shown that detecting the dephasing in the signal of such a system is in principle possible, but only realistic for very massive ($\sim$10$^6$ M$_{\sun}$) and close (sub-parsec) perturbers.

The shift in energy is not the only way in which a third body can perturb the hard binary. For orbits that are sufficiently misaligned, Kozai--Lidov (KL; \citealt{Kozai,Lidov}) oscillations can induce a secular variation in the eccentricity of the orbit. If the inspiral time-scale is longer than the time-scale of KL oscillations, $t_{\rm{KL}}$, the evolution equations for the orbital parameters of the hard binary can be significantly affected by the perturber. The detectability of KL oscillations in the GW signal of LISA sources has been very recently studied in \cite{2019arXiv190208604R} and \cite{2020ApJ...901..125D}, where the original version of Peters' time-scale is used as a simple model for the inspiral time-scale. An intermediate calculation in the aforementioned papers is the computation of the characteristic orbital separation, $a_{\rm{KLO/GW}}$, of the inner binary at which KL oscillations are quenched by GW emission. The radius is found by equating the GW and KL time-scales and solving for the semi-major axis:

\begin{align}
     &t_{\rm{GW}} \sim R(e) Q_{\rm{h}}(a,e)Q_{\rm{s}}(a,e,s_1) t_{\rm P} \nonumber
     = t_{\rm{KL}} \\ &= 2 \pi \frac{\sqrt{G M}}{G m_3}\frac{R^3}{a^{3/2}} \\
     &\implies a_{\rm{KLO/GW}} \approx \nonumber\\ & 2 \left(\frac{2^7 \pi M q R^3 f(e) (G M)^{5/2}}{5c^5 m_3
   R(e) Q_{\rm{h}}(a,e)Q_{\rm{s}}(a,e,s_1)}\right)^{2/11},\\ \nonumber
\end{align}

\noindent where we model the corrections to Peters' time-scale with the correction factors $R(e) Q_{\rm{h}}(a,e)Q_{\rm{s}}(a,e,s_1)$. The general solution $a_{\rm{KLO/GW}}$ must be computed numerically. However, we can see that (while suppressed by the small power) the correction factors can reduce the value of the quenching radius $a_{\rm{KLO/GW}}$ by a factor of up to $\sim$1.5, which might be enough to significantly affect any further results that depend on powers of $a_{\rm{KLO/GW}}$. An example would be the phase-space volume $a_{\rm{KLO/GW}}^3$, which can easily change by a factor of 2--3.

\subsection{Gas drag and torques}

Gas drag and torques in accretion discs are expected to significantly influence the dynamics of an inspiral event and might cause a detectable dephasing of the GW signal depending on the physical properties of the disc \citep[see, e.g.][]{2014PhRvD..89j4059B, 2019MNRAS.486.2754D}. While several different disc models exist, many show a central region that extends for $\sim$$5\times 10$$^3$ Schwarzschild radii, where the density and temperature profiles are much more shallow than in the external regions \citep[see the profiles in, e.g.][]{1973A&A....24..337S,2003MNRAS.341..501S,2009ASPC..408..128T}. If we restrict our analysis to orbits within this region, it is a reasonable first-order approximation to describe the local density and temperature as constant. To describe the influence of the disc on the inspiral time-scale, we need to model the effects that have a direct influence on the orbital elements of the secondary BH.

One such effect is gas drag. While the subject of dynamical friction is complex, we will assume that, for BHs, gas drag can be locally modelled by the Bondi--Hoyle--Lyttleton (BHL) drag \citep[see, e.g.][]{1939PCPS...35..592H,1944MNRAS.104..273B,1952MNRAS.112..195B,1999ApJ...513..252O,2020MNRAS.tmp.2858F}. At every completed orbit, the energy of the secondary BH will be reduced by an amount of work $W_{\rm{BHL}}$ given by

\begin{align}
    W_{\rm{BHL}}&= \int_{\mathcal{S}} \mathbf{F_{\rm{BHL}}} d\mathbf{s},\\ \nonumber
\end{align}

\noindent where $\mathcal{S}$ is the path of the orbit in space. The magnitude of the BHL drag reads

\begin{align}
F_{\rm{BHL}}= \frac{4\pi G^2 m^2 \rho}{c_{\rm s}^2 + v^2_{\rm{rel}}},
\end{align}

\noindent where $c_{\rm s}$ is the local speed of sound, $\rho$ is the local disc density, $m$ is the mass of the inspiralling BH, and $v_{\rm{rel}}$ the relative velocity with respect to the Keplerian disc velocity. If the orbit is very eccentric, the instantaneous velocity, $v_{\rm i}$, of the secondary will generally be misaligned with the Keplerian disc velocity. In this case we can approximate $v_{\rm{rel}} \approx v_{\rm i}$ and $v_{\rm i} \gg c_{\rm s}$, which reduces the BHL drag formula to the standard high-velocity limit of dynamical friction \citep[see, e.g.][for a discussion of gas effects in highly eccentric orbits]{2013CQGra..30x4008M}. Note that this might also be the more appropriate description, at least in a local sense, in discs that are very turbulent, where gas velocity deviates from a Keplerian disc (but only if the sound speed is also smaller than the BH velocity). We will consider both the BHL case and the high-velocity dynamical friction case, keeping in mind that the former is more accurate for circular orbits and the latter is more accurate for eccentric ones. The integral can only be solved numerically or as a series expansion in the eccentricity. We compute it for the two cases:

\begin{align}
    \label{eq:work}
   W_{\rm{BHL}}  &= \frac{8 a G^2 M^2 q^2 \pi^2 \rho}{c_{\rm s}^2} \mathcal{W}_{\rm{BHL}}(e,n_{\rm M}), \\
   W_{\rm{dyn}} & = 8 a^2 G M q^2 \pi^2 \rho \mathcal{W}_{\rm{dyn}}(e),\\ \nonumber
\end{align}

\noindent where $\mathcal{W}_{\rm{BHL}}$ is a function that depends on the eccentricity of the orbit and an effective average Mach number $n_{\rm M} = v_{\rm d}^2c_{\rm s}^{-2} $ evaluated at a radius of one semi-major axis, where $v_{\rm d}$ is the Keplerian disc velocity. It varies between one (for circular orbits) and several thousands, depending on the eccentricity, and converges very poorly for high Mach numbers. The function $\mathcal{W}_{\rm{dyn}}$ only depends on eccentricity and can be computed as a polynomial expression in $e^2$:

\begin{align}
    \mathcal{W}_{\rm{dyn}}(e) = 1 +\frac{3}{4}e^2 + \frac{21}{64}e^4+ \frac{55}{256}e^6 + \mathcal{O}[10^{-1}e^8].\\ \nonumber
\end{align}

The loss of energy due to drag results in a secular change of the semi-major axis of the secondary BH, which we can compare to similar effects caused by the PN fluxes. To obtain an expression for the orbit-averaged energy flux due to BHL drag and dynamical friction, we simply divide Equation~\eqref{eq:work} by an orbital period $T$.

Keeping this result in mind, we turn our attention to another effect that directly influences the orbit of the secondary BH, namely global torques. A realistic description of torques in accretion discs is only possible through hydro-dynamical simulations. Nevertheless, we will assume that the simplest analytical models of Type~I and Type~II viscous torques are an appropriate order-of-magnitude approximation, when averaged over many orbits. We take the formulas by \cite{2002ApJ...565.1257T} and \citet{2020arXiv200511333D}, which read

\begin{align}
    \dot{L}_{\rm I} &=T^{-2} \rho h r(t)^4 q^2 \left(\frac{v_{\rm d}}{c_{\rm s}}\right)^2 =  \rho h r(t)^3 \frac{G^2 m^2}{4 \pi a^3 c_{\rm s}^2}, \\
    \dot{L}_{\rm {II}}&= \alpha 3 \pi r(t)^2 \Omega _2 c_{\rm s} \rho h^2,\\ \nonumber
\end{align}

\noindent where $\Omega_2$ is the Keplerian frequency of the secondary and $\alpha$ is the viscous parameter, and we assume that the disc scale height, $h$, follows the standard scaling $h\sim c_{\rm s} \Omega ^{-1}$, where $\Omega$ is the Keplerian disc orbital frequency. We average the expressions over one orbit, yielding

\begin{align}
\left< \dot{L}_{\rm 
i} \right>  &= \frac{1}{T} \int_0^T\dot{L}_{\rm i} \dot{\phi}^{-1} d\phi \\
\implies \left< \dot{L}_{\rm I} \right> &= \frac{\sqrt{a^{3} G^{3} M^{3}} q^2 \rho }{4 \pi ^{2} c_{\rm s}}\mathcal{D}_{\rm I}(e)  \label{eq:typeI} \\
\implies \left< \dot{L}_{\rm II} \right> &=  \alpha  \frac{3}{2}\frac{ a^{7/2} c_{\rm s}^3 \rho }{\sqrt{G M}} \mathcal{D}_{\rm II}(e)  \label{eq:typeII}\\ \nonumber
\end{align}

\noindent where $\mathcal{D}_{\rm i}(e)$ are functions of the eccentricity. The function $\mathcal{D}_{\rm I}(e)$ is approximated very well by the first four orders of its polynomial expansion, whereas $\mathcal{D}_{\rm {II}}(e)$ has a closed form solution:

\begin{align}
    \mathcal{D}_{\rm I}(e) &=1 + \frac{99 }{16}e^2 + \frac{3465 }{1024}e^4 + \frac{1155}{16384}e^6+ \mathcal{O}[10^{-3}  e^8], \\
    \mathcal{D}_{\rm II}(e) &=1+\frac{15}{2} e^2+\frac{45 }{8}e^4+\frac{5}{16}e^6.\\ \nonumber
\end{align}

In the case of torques, the orbit-averaging procedure is not completely legitimate: Type~I torques are a consequence of resonances between the orbital frequency of the secondary and the disc, and as such it is unlikely that they would realistically appear for eccentric orbits. Type~II torques become relevant when the secondary is massive enough to open gaps in the disc structure \citep[see, e.g.][]{1980ApJ...241..425G,1997Icar..126..261W}. In a tight binary of SMBHs, the gap opening torques result into the formation of a cavity, and of a circumbinary disc around it \citep[e.g.][]{cuadra,2012A&A...545A.127R,2017MNRAS.469.4258T}. These torques depend on the detailed thermodynamics and viscous properties of the disc because both pressure forces and viscous forces oppose the gap-opening torque induced by the binary into the surrounding disc gas. Therefore, one would in principle need to account for the dependence of the torques on the many disc properties that determine the strength of the forces at play. However, here we take an explorative stand as we only want to highlight order-of-magnitude effects, which could later be explored in a self-consistent model for the disc. In general, Type~II torques will apply only to SMBH binaries as opposed to IMRIs/EMRIs, because there must be a massive enough secondary to open a gap. The recent sub-pc scale hydrodynamical simulations of \citet{2020ApJ...899..126S} show that even with a mass ratio of 20:1 the opening of a gap or cavity, is not always occurring, whereas \citet{2020arXiv200511333D} show that, for IMRIs, the orbital evolution proceeds with no gap opening. Additionally, we will also see that the functions $\mathcal{D}_{\rm i}$ do not play a significant role for the results of this section.

Equations~\eqref{eq:typeI} and \eqref{eq:typeII} are in the form of an orbit-averaged angular momentum flux, which we can compare directly to the PN fluxes.\footnote{In the following calculations, we will always assume that the spin of the primary BH does not evolve due to gas accretion, at least over the time-scale of a single inspiral event. In this case, we can use the vacuum solutions that were discussed in the previous sections.} Note that, in the case of circular orbits, one can obtain results that are consistent with any particular disc model by simply replacing $\rho$ and $c_{\rm s}$ with the corresponding profiles. For the eccentric case, the orbit averages can be computed numerically for arbitrary disc models.

In order to compare a large range of PN orders to the results shown above, we have to find a simple way to characterise their scaling in the semi-major axis and eccentricity of the secondary's orbit. By checking up to the first four integer-PN orders of both energy and angular momentum fluxes, we always find the following structure:

\begin{align}
\dot{E}_{\rm{n-PN}} &\propto \dot{E}_{\rm{N}} \left( \frac{r_{\rm S}}{a}\right)^n \left( \frac{C_{\rm{n}}(e)}{(1-e^2)^n}\right), \\
\dot{L}_{\rm{n-PN}} &\propto \dot{L}_{\rm{N}} \left( \frac{r_{\rm S}}{a}\right)^n \left( \frac{D_{\rm{n}}(e)}{(1-e^2)^n}\right),\\ \nonumber
\end{align}

\noindent where the subscript $n$ denotes the (integer) PN order and the subscript $N$ denotes the lowest-order (quadrupolar) fluxes, whereas the coefficients $C_{i}$ and $D_{i}$ are fractions of polynomials in $e^2$ of order unity. The half-integer terms have slightly different forms, but we will neglect them fur the purpose of the following calculations.

To find a series of characteristic orbital separations, at which the effect of the gas on the evolution of the orbit is as strong as the $n$-th order PN fluxes, we equate the expressions and solve for the semi-major axis. Note that this is equivalent to equating the GW-induced inspiral time-scale to a characteristic inspiral time-scale due to the effect of the gas. In the case of BHL drag, we can solve the equation analytically only for the case of circular orbits, because in general the function $\mathcal{W}_{\rm{BHL}}$ contains an arbitrary amount of powers of the semi-major axis $a$. For circular orbits, the characteristic semi-major axis $a_{\rm{BHL/PN}}$ reads

\begin{align}
a_{\rm{BHL/PN}} ^{e\to 0}& \approx r_{\rm S} \left(\frac{c_{\rm s}^2 }{G r_{\rm S}^2 \rho} \right)^{\frac{2}{9+2n}}, \label{eq:r10}\\ \nonumber
\end{align}

\noindent where we are able to neglect all the small numerical coefficients because of the small exponent. For dynamical friction, the radii $a_{\rm{dyn/PN}}$ read

\begin{align}
a_{\rm{dyn/PN}} & \approx r_{\rm S} \left(\frac{c^2 }{G r_{\rm S}^2 \rho} \right)^{\frac{2}{11+2n}}\left( \frac{f(e)(1-e^2)^{-n}}{\mathcal{W}_{\rm{dyn}}(e)} \right)^{\frac{2}{11+2n}},\\
a_{\rm{dyn/PN}} & \approx r_{\rm S} \left(\frac{c^2 }{G r_{\rm S}^2 \rho} \right)^{\frac{2}{11+2n}}\left( 1-e^2\right)^{-\frac{7+2n}{11+2n}},\\ \nonumber
\end{align}

\noindent where, in the last step, we approximated $[f(e)/ \mathcal{W}_{\rm{dyn}}(e)]^{2/11} \simeq 1$. We will use the BHL results as the low-eccentricity limit and the dynamical friction results as the high-eccentricity limit.

We repeat the same calculation for the case of Type~I torques and obtain similar characteristic separations $a_{\rm{TI/PN}}$ as before. In this case, we are able to solve it analytically for arbitrary eccentricities, with the caveat that global torques are unlikely to appear for non-circular orbits:

\begin{align}
  a_{\rm{TI/PN}}  &\approx  r_{\rm S} \left(\frac{c_{\rm s} c }{ G  r_{\rm S}^2 \rho} \right)^{\frac{1}{5+n}} \left(\frac{(1-e^2)^{-2 -n}}{\mathcal{D}_{\rm I}(e)} \right) ^{\frac{1}{5+n}} \\
  &\approx  r_{\rm S} \left(\frac{c_{\rm s} c }{ G  r_{\rm S}^2 \rho} \right)^{\frac{1}{5+n}} \left(1-e^2\right) ^{-\frac{2 + n}{5+n}}.\label{eq:TIPN}\\ \nonumber
\end{align}

Here we have a slightly different scaling in the physical properties of the disc, but otherwise the formulas look very similar. Equation~\eqref{eq:TIPN} is a good approximation even for the whole eccentricity range, because of the small range in values of the function $\mathcal{D}_{\rm I}(e)$ in the interval $e \in [0,1]$. The same is true for Type~II torques, where the characteristic radii $a_{\rm{TII/PN}}$ read

\begin{align}
  a_{\rm{TII/PN}}  &\approx  r_{\rm S} \left(\frac{ c^5 q ^2 }{\alpha c_{\rm s}^3 G  r_{\rm S}^2 \rho} \right)^{\frac{1}{7+n}} \left(\frac{(1-e^2)^{-2 -n}}{\mathcal{D}_{\rm II}(e)} \right) ^{\frac{1}{7+n}} \\
  &\approx  r_{\rm S}\left(\frac{ c^5 q^2}{\alpha c_{\rm s}^3 G  r_{\rm S}^2 \rho} \right)^{\frac{1}{7+n}} \left(1-e^2\right) ^{-\frac{2 + n}{7+n}}. \label{eq:r11}\\ \nonumber
\end{align}

With the aid of these characteristic radii, we can identify different regions of phase space in which we can expect certain PN orders to be accurate descriptions of the inspiral time-scale. The Newtonian ($n=0$) result is particularly telling, since it separates the regions where gas effects dominate from regions where GW emission dominates.

\begin{figure}
    \centering
    \includegraphics[scale=0.7]{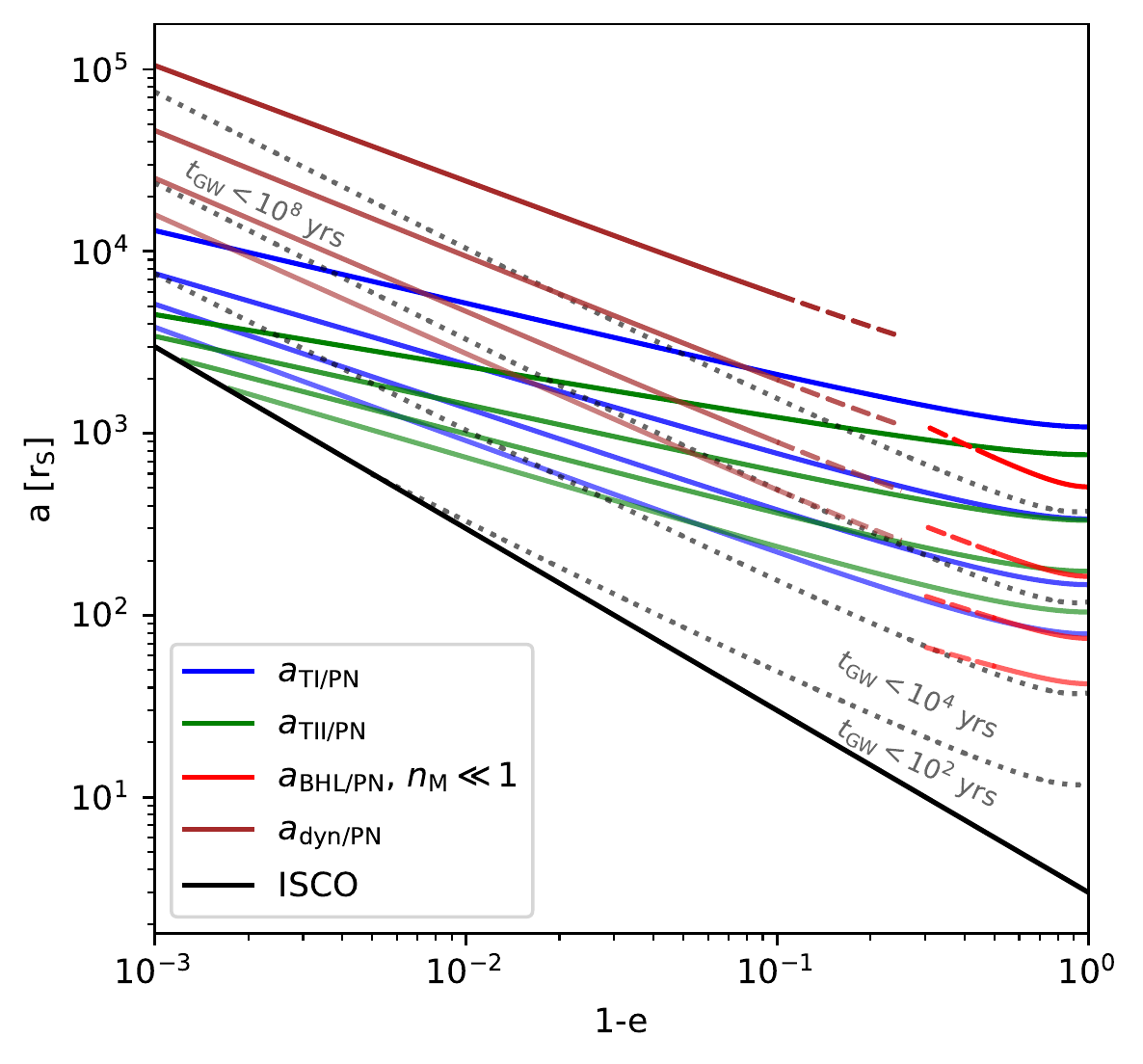}
    \caption{Shown is a region of phase space in which the characteristic separations $a_{\rm{TI/PN}}$, $a_{\rm{TII/PN}}$, $a_{\rm{BHL/PN}}$, and $a_{\rm{dyn/PN}}$ are plotted. From the highest to the lowest, curves of the same colour separate regions of phase space in which the effect of gas is stronger than the 0th, first, second, and third PN order effects. In other words, one can expect this particular GW source to decay in a time that is well described by $n$-th order PN theory only if it starts from a location \textit{below} the $n$-th set of coloured lines in phase space. The results shown are for an SMBH of $10^6$~M$_{\sun}$ and a secondary of $10$~M$_{\sun}$ embedded in an accretion disc with a central mean density of $\rho=10^{-13}$~g~cm$^{-3}$ and a viscous coefficient $\alpha$ of $10^{-3}$, consistent with the models in \citet{2009ASPC..408..128T} . Changing the density and viscosity parameter of the central part of the disc by orders of magnitude only slightly shifts the coloured curves up and down. The black line marks the location of the ISCO for a non-rotating central object, and the dotted lines denote the isochrone curves for different GW decay times.}
    \label{fig:gas}
\end{figure}

Figure~\ref{fig:gas} shows this procedure for the standard case of a $10^6$ M$_{\sun}$ SMBH and a $10$ M$_{\sun}$ compact object. The results only scale very weakly with disc properties such as central density and temperature, suggesting that they are robust as an order-of-magnitude estimate. For example, one can use the formulas for the characteristic radii to estimate deviations from the vacuum solutions as a function of the disc properties. As an example, let us take a $10$ M$_{\sun}$ BH on a circular orbit at a radius of $10^3$ $r_{\rm S}$. If it is embedded in a disc with a local density of $\rho=10^{-13}$ g cm$^{-3}$ and viscosity parameter $\alpha = 10^{-3}$  \citep{2009ASPC..408..128T}, we expect the effect of BHL drag (if present) to be almost as important as the lowest-order quadrupole radiation, while Type~I or Type~II torques (if present) would only be as important as a 1~PN correction.

\begin{figure}
    \centering
    \includegraphics[scale=0.7]{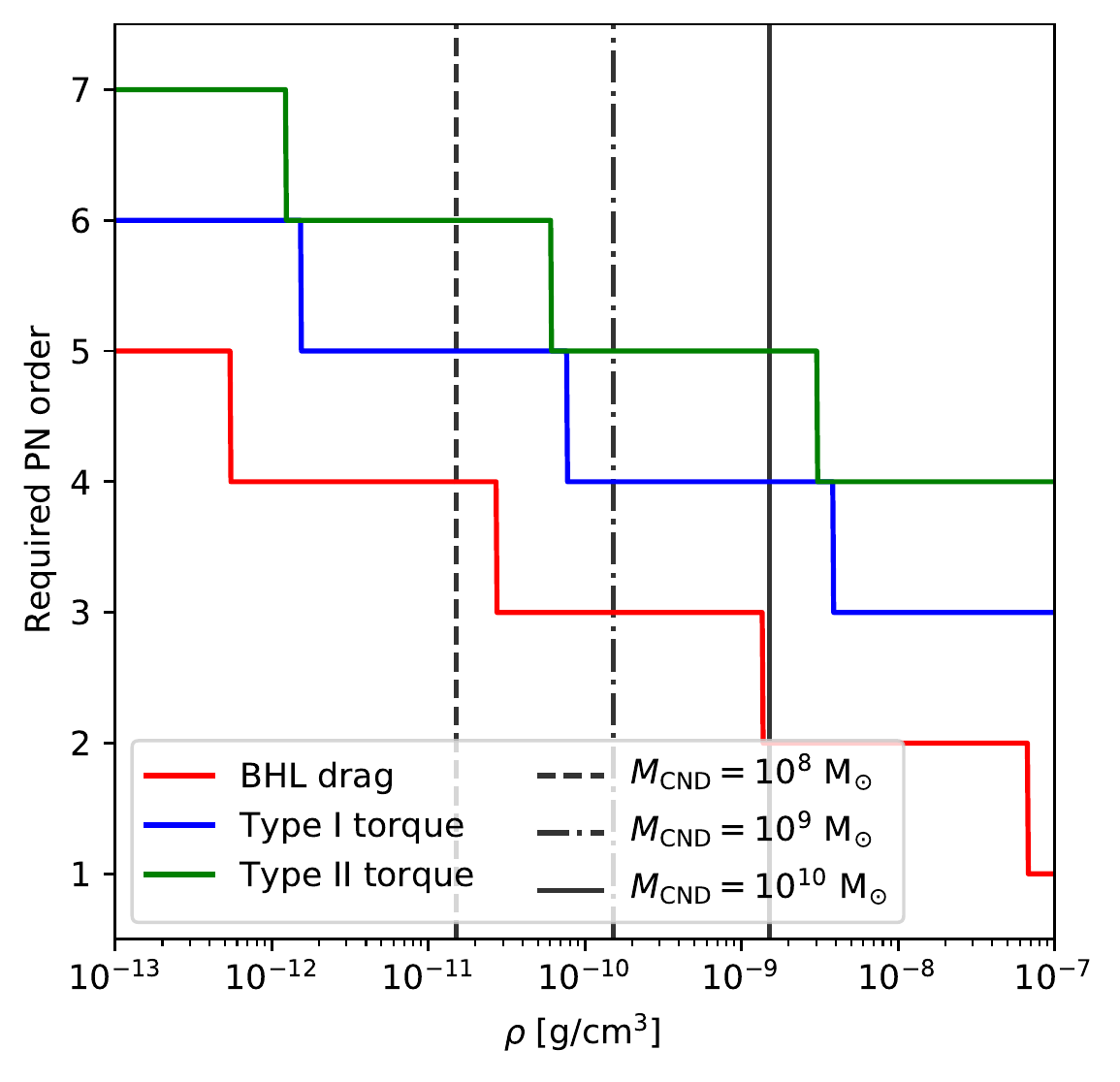}
    \caption{We show the required PN order needed to detect a shift from the vacuum time-scale in the cases of BHL drag, Type~I and Type~II torques for different values of the local density. The lines are computed for the standard case of a $10^6$ M$_{\odot}$ SMBH and a 10 M$_{\odot}$ secondary entering the LISA band at $\sim$1~mHz (solid coloured lines) with a local sound speed of $\sim$ $3 \times 10^7$~cm~s$^{-1}$. The vertical lines represent a qualitative extrapolation of the local density from observed CND masses and extensions. The results only scale very weakly with parameters such as the CND mass, local sound speed, or the mass ratio of the binary, but change significantly with the SMBH mass. This is, however, not due to the properties of the disc, but rather to the fact that the efficiency of GW emission at a fixed frequency strongly depends on the mass of the SMBH.}
    \label{fig:gaspn}
\end{figure}

Vice versa, one can also estimate the disc properties required for a shift to be visible at a certain PN order, which we show in Figure \ref{fig:gaspn} for varying values of the local disc density. One can also use this type of analysis to speculate on the likely environment of actual LISA sources. As an example, if we assume that the density of a typical circumnuclear disc (CND) scales as a simple $r^{-2}$ power law \citep{1963MNRAS.126..553M} until it flattens out at a radius of $\sim$$5 \times 10^3 $$r_{\rm S}$, we can extrapolate some realistic densities from observational results for the total mass and extension of CNDs given in \cite{2014ApJ...784...70M}, obtaining $\rho \sim 10^{-11}$ to $10^{-9}$~g~cm$^{-3}$ \citep[a result more consistent with the disc models in][]{2003MNRAS.341..501S}. Note that these assumptions are also used as initial conditions in state-of-the-art simulations of SMBH coalescence \citep[see, e.g.][]{2017ApJ...838...13S,2020ApJ...899..126S}. Even for the wide range of observed CND masses ($\sim$$10^8$~M$_{\sun}$ to $\sim$$5\times 10^9$~M$_{\sun}$), the required PN order to detect a shift in the inspiral time-scale for an EMRI that is entering the LISA band (at $\sim$1~mHz) only ranges between 3 and 5, which is realistic given a source with sufficient SNR. These results also qualitatively agree with more sophisticated hydrodynamic simulations such as those in \citet{2019MNRAS.486.2754D} and \citet{2020arXiv200511333D}, where it is shown that, if surface densities of $ \gsim 10^3$ g cm$^{-2}$ are present, a dephasing of the GWs is likely to be detectable by LISA. Note that this type of calculation can be repeated with more sophisticated density profiles to yield more precise results. However, even very simple estimates such as assuming a $r^{-2}$ density profile until $5 \times 10^3$ $r_{\rm S}$ can yield correct order-of-magnitude estimates because of the weak scaling of the characteristic radii with disc properties.

\section{Summary and Conclusion}\label{sec:conclusions}

In this paper, we extended our analysis of the GW-induced decay time-scale to include the 1.5~PN order. We produced a series of analytical fits that can be multiplied by Peters' formula in order to model hereditary and spin-orbit effects (Equations~\ref{eq:r1} to \ref{eq:r2}). In the high-eccentricity (Equation~\ref{eq:r4}) and low-eccentricity (Equation~\ref{eq:r3}) limits, the fits reduce to simple exponential functions that can be used to avoid errors of a factor $\sim$10 and $\sim$2, respectively. We show that these modifications to Peters' formula are necessary in applications that are highly sensitive to the inspiral time-scale, such as the calculation of the EMRI critical semi-major axis (\citealt{pau_2013,Amaro_Seoane_2019}; see also V\'{a}zquez-Aceves et al., in prep.).

We then used the results of \cite{Will_KL} in combination with our findings to compare the strength of arbitrarily high PN perturbations against gravitational perturbations in a hierarchical triple system. We found a series of characteristic radii at which the perturbations affect the inspiral time-scale as much as any particular PN order (Equation~\ref{eq:rpn}), which we used to reproduce previously known results on the detectability of the dephasing of GW signals \citep{PhysRevD.83.044030}. Furthermore, we briefly discussed the effects of KL oscillation and how corrections to Peters' time-scale can affect recent calculations found in \cite{2019arXiv190208604R} and \cite{2020ApJ...901..125D} that depend on the typical radius at which KL oscillations are quenched by GW emission.

Finally, we discussed the validity of vacuum time-scales for inspirals embedded in a gaseous medium, which is the likely case for SMBHs sourrounded by accretion discs and CNDs. We analysed several types of drag forces (BHL drag and dynamical drag) and global torques (Type~I torques and Type~II viscous torques), and compared them against PN energy and angular momentum fluxes of various orders. We found several characteristic radii that separate the phase space in different dynamical regions, where gas effects are as strong as a particular PN order (Equations~\ref{eq:r10} to \ref{eq:r11}). We used a simple prescription to extrapolate the gas densities that would be likely encountered by a GW source entering the LISA band in one of the several observed CNDs \citep[][]{2014ApJ...784...70M}, and found that PN orders of 3--5 are required to detect a deviation from the vacuum case, which is compatible with results from hydrodynamical simulations \citep{2019MNRAS.486.2754D,2020arXiv200511333D}. While many assumptions are required to derive the analytical formulas, the results scale only very weakly with disc properties, suggesting that they might be a robust order-of-magnitude approximation of a more realistic treatment of gas-embedded inspirals.

\section*{Acknowledgements}
The authors thank the anonymous reviewer for providing feedback that greatly improved this work. EB, PRC, LM, and LZ acknowledge support from the Swiss National Science Foundation under the Grant 200020\_178949. EB  acknowledges support from the European Research Council (ERC) under the European Union's Horizon 2020 research and innovation program ERC-2018-COG
under grant agreement N.~818691 (B~Massive). PAS acknowledges support from the Ram{\'o}n y Cajal Programme of the Ministry of Economy, Industry and Competitiveness of Spain, the COST Action GWverse CA16104, the National Key R\&D Program of China (2016YFA0400702) and the National Science Foundation of China (11721303).

\section*{Data Availability Statement}
The data underlying this article will be shared on reasonable request to the corresponding author.

\scalefont{0.94}
\setlength{\bibhang}{1.6em}
\setlength\labelwidth{0.0em}
\bibliographystyle{mnras}
\bibliography{references}
\normalsize

\appendix

\section{Parametrisation and Evolution Equations}\label{sec:appendix}

In this appendix, we list the PN parametrisation and evolution equations that were necessary for this work, adjusted to our notation.

\subsection*{First Order and Hereditary terms}

From \cite{2019PhRvD.100h4043E}, we take the 1~PN fluxes and 1.5~PN hereditary fluxes radiation reaction equations:

\begin{align}
	\frac{d x}{d t} &= \frac{2 x^5}{3} \left(\mathcal{X}_{\rm N}+ x \mathcal{X}_{\rm{1PN}}  +x^{3/2}\mathcal{X}_{\rm{tail}}\right),\\
	\frac{d e_{\rm d}}{d t} &= x^4 \left(\mathcal{E}_{\rm N}+ x \mathcal{E}_{\rm{1PN}}  +x^{3/2}\mathcal{E}_{\rm{tail}}\right),\\ \nonumber
\end{align}

\noindent where:

\begin{align}
    \mathcal{X}_{\rm N} &= \frac{1}{\left(1-e_{\rm d}^2\right)^{7/2}} \left\{ \frac{96}{5}+\frac{292 
		e_{\rm d}^2}{5} +\frac{37 e_{\rm d}^4}{5}\right\},\\
	\mathcal{X}_{\rm{1PN}} &= \frac{1}{\left(1-e_{\rm d}^2\right)^{9/2}}\Bigg\{ -\frac{1486}{35} 
	-\frac{264 \nu}{5} \nonumber \\ &+e_{\rm d}^2 \left(\frac{2193}{7}-570 \nu \right) +e_{\rm d}^4 
		\left(\frac{12217}{20}-\frac{5061 \nu}{10}\right) \nonumber \\ &+e_{\rm d}^6 \left(\frac{11717}{280}-\frac{148 
		\nu}{5}\right) \Bigg\}, \\
	\mathcal{X}_{\rm{tail}}&=x^{3/2}\frac{96}{5}\phi (e_{\rm d}),\\
	\phi (e_{\rm d}) &=  1+\frac{2335}{192} e_{\rm d}^2 +\frac{42955}{768} e_{\rm d}^4 +\frac{6204647}{36864} e_{\rm d}^6 + \mathcal{O}[e_{\rm d}^8],\\ \nonumber
	\end{align}
	
\noindent and
	
	\begin{align}
	\mathcal{E}_{\rm N} &= \frac{1}{\left(1-e_{\rm d}^2\right)^{5/2}} \left\{ \frac{304}{15}+\frac{121 
		e_{\rm d}^2}{15} \right\},\\
	\mathcal{E}_{\rm{1PN}} &= \frac{1}{\left(1-e_{\rm d}^2\right)^{7/2}}\Bigg\{ -\frac{939}{35} 
		-\frac{4084 \nu}{45}\nonumber +e_{\rm d}^2 \left(\frac{29917}{105} - \frac{7753}{30} \nu \right)\nonumber \\ &+e_{\rm d}^4 
		\left(\frac{13929}{280} - \frac{1664 \nu}{45}\right) \Bigg\},\\
	\mathcal{E}_{\rm{tail}}&=-\frac{32}{5} \left(-\frac{985}{48} \pi  x^{3/2} \psi(e_{\rm{d}})\right)\\
		\psi (e_{\rm d}) &=  \frac{192}{985} \frac{\sqrt{1-e_{\rm{d}}^2}}{e_{\rm{d}}^2} \left[\sqrt{1-e_{\rm{d}}^2} \phi(e_{\rm{d}}) 
		-\tilde{\phi}(e_{\rm{d}})\right],\\
	\tilde{\phi}(e_{\rm{d}})&=1+\frac{209}{32}  e_{\rm{d}}^2+\frac{2415}{128} e_{\rm{d}}^4 +\frac{730751}{18432} 
		e_{\rm{d}}^6+ \mathcal{O}[e_{\rm d}^8].\\ \nonumber
\end{align}

\subsection*{Spin-Orbit Coupling}

From \cite{jetzer}, we take the 1.5~PN spin-orbit coupling parametrization equations,

\begin{align}
a &= x^{-1} \left[ 1 + \frac{x^{3/2} \beta(2,1)}{\sqrt{1-e^2}} 
\right], \\
e &= e_{\rm d} \left[ 1 - \frac{x^{3/2} \beta \left(2, 1
\right)}{\sqrt{1-e_{\rm d}^2}} \right] \\ \nonumber
\end{align}

\noindent as well as the 1.5 PN spin-orbit coupling fluxes equations:

\begin{align}
\frac{dn}{dt} &= \frac{\nu x^{11/2}}{\left(1 - e_{\rm d}^2\right)^{7/2}} \Bigg[
\frac{1}{5} \left( 96 + 292 e_{\rm d}^2 + 37 e_{\rm d}^4 \right) \nonumber\\
&-
\frac{x^{3/2}}{10\left(1 - e_{\rm d}^2\right)^{3/2}} \beta ( 3088 +
\numprint{15528}e_{\rm d}^2 + 7026 e_{\rm d}^4 \nonumber \\
&+ 195e_{\rm d}^6, 2160 + \numprint{11720} e_{\rm d}^2 + 5982 e_{\rm d}^4
+ 207 e_{\rm d}^6 )\Bigg], \\
\frac{de_{\rm d}^2}{dt} &= -\frac{\nu x^4}{\left(1 - e_{\rm d}^2\right)^{5/2}} \Bigg[
\frac{2e_{\rm d}^2}{15} \left( 304 + 121 e_{\rm d}^2 \right) \nonumber\\
&- \frac{e_{\rm d}^2 x^{3/2}}{15\left(1 -
e_{\rm d}^2\right)^{3/2}} \beta ( \numprint{13048} + \numprint{12000} e_{\rm d}^2 + 789
e_{\rm d}^4, \nonumber \\
& 9208 + \numprint{10026} e_{\rm d}^2 + 835 e_{\rm d}^4 )\Bigg],\\ \nonumber
\end{align}

\noindent where

\begin{align}
n &= x^{3/2},\\
\beta(a,b) &= \frac{\mathbf{J}}{J} \cdot \left( a \mathbf{\zeta} + b \mathbf{ \xi } \right),\\
 \mathbf{ \zeta } &= \mathbf{S}_1 + \mathbf{S}_2, \\
 \mathbf{\xi} &= \frac{m_2}{m_1} \mathbf{S}_1 + \frac{m_1}{m_2} \mathbf{S}_2. \\ \nonumber
 \end{align}


\bsp 
\label{lastpage}
\end{document}